\documentclass[twocolumn]{aastex63}

\begin{document}
 
\newcommand{\kms}{km s$^{-1}\;$}
\newcommand{\msun}{M_{\odot}}
\newcommand{\rsun}{R_{\odot}}
\newcommand{\lsun}{L_{\odot}}
\newcommand{\teff}{T_{\rm eff}}
\newcommand{\kep}{{\it Kepler}~}
% Roman Numerals
\makeatletter
\newcommand{\Rmnum}[1]{\expandafter\@slowromancap\romannumeral #1@}
\newcommand{\rmnum}[1]{\romannumeral #1}
 
\title{The K2 M67 Study: Precise Mass for a Turnoff Star in the Old Open Cluster M67}

\author[0000-0003-4070-4881]{Eric L. Sandquist}
\affiliation{San Diego State University, Department of Astronomy, San
  Diego, CA, 92182 USA}

\author[0000-0001-9911-7388]{David W. Latham}
\affiliation{Center for Astrophysics | Harvard \& Smithsonian, 60 Garden Street,
  Cambridge, MA 02138, USA}

\author[0000-0002-7130-2757]{Robert D. Mathieu}
\affiliation{University of Wisconsin-Madison, Department of
  Astronomy, Madison, WI 53706, USA}

\author[0000-0002-3944-8406]{Emily Leiner}
\affiliation{Center for Interdisciplinary Exploration and Research in Astrophysics,
Northwestern University, Evanston, IL 60208, USA}
\affiliation{University of Wisconsin-Madison, Department of
  Astronomy, Madison, WI 53706, USA}
\affiliation{NSF Astronomy and Astrophysics Postdoctoral Fellow}

\author[0000-0001-7246-5438]{Andrew Vanderburg}
\affiliation{Harvard-Smithsonian Center for Astrophysics,
  Cambridge, MA 02138, USA}

\author[0000-0002-4879-3519]{Dennis Stello}
\affiliation{School of Physics, The University of New South Wales,
  Syndey NSW 2052, Australia}
\affiliation{Sydney Institute for Astronomy (SIfA), School of Physics,
  University of Sydney, NSW, 2006, Australia}
\affiliation{Stellar Astrophysics Centre, Department of Physics and
  Astronomy, Aarhus University, Ny Munkegade 120, DK-8000 Aarhus C,
  Denmark}

\author{Jerome A. Orosz} 
\affiliation{San Diego State University, Department of Astronomy, San
  Diego, CA, 92182 USA}

\author{Luigi R. Bedin}
\affiliation{Istituto Nazionale Astrofisica di Padova -
  Osservatorio Astronomico di Padova, Vicolo dell'Osservatorio 5,
  I-35122 Padova, Italy}

\author[0000-0001-9673-7397]{Mattia Libralato}
\affiliation{Space Telescope Science
  Institute, 3700 San Martin Drive, Baltimore, MD 21218, USA}
\affiliation{Istituto Nazionale Astrofisica di Padova -
  Osservatorio Astronomico di Padova, Vicolo dell'Osservatorio 5,
  I-35122 Padova, Italy}
\affiliation{Dipartimento di Fisica e Astronomia `Galileo
  Galilei', Universit\`{a} di Padova, Vicolo dell'Osservatorio 3,
  Padova I-35122, Italy}
  
\author{Luca Malavolta}
\affiliation{Istituto Nazionale Astrofisica di Padova -
  Osservatorio Astronomico di Padova, Vicolo dell'Osservatorio 5,
  I-35122 Padova, Italy}
\affiliation{Dipartimento di Fisica e Astronomia `Galileo
  Galilei', Universit\`{a} di Padova, Vicolo dell'Osservatorio 3,
  Padova I-35122, Italy}

\author{Domenico Nardiello}
\affiliation{Aix Marseille Univ, CNRS, CNES, LAM, Marseille, France}
\affiliation{Istituto Nazionale Astrofisica di Padova -
  Osservatorio Astronomico di Padova, Vicolo dell'Osservatorio 5,
  I-35122 Padova, Italy}

\correspondingauthor{Eric L. Sandquist}
\email{esandquist@sdsu.edu}

\begin{abstract}
  We present a study of the bright detached eclipsing main sequence
  binary WOCS 11028 (Sanders 617) in the open cluster M67. Although
  the binary has only one eclipse per orbital cycle, we show that the
  masses of the stars can be derived very precisely thanks to a strong
  constraint on the orbital inclination: $M_A = 1.222\pm0.006 \msun$
  and $M_B = 0.909\pm0.004 \msun$. We use a spectral energy
  distribution fitting method to derive characteristics of the
  component stars in lieu of the precise radii that would normally be
  derived from a doubly-eclipsing binary.  The deconvolution of the
  SEDs reveals that the brighter component of the binary is at the
  faint turnoff point for the cluster --- a distinct evolutionary
  point that occurs after the convective core has been established and
  while the star is in the middle of its movement toward lower surface
  temperature, before the so-called hook at the end of main
  sequence. The measurements are in distinct disagreement with
  evolution models at solar metallicity: higher metal abundances are
  needed to reproduce the characteristics of WOCS 11028 A.
We discuss
    the changes to model physics that are likely to be needed to
    address the discrepancies. The clearest conclusions are that
    diffusion is probably necessary to reconcile spectroscopic
    abundances of M67 stars with the need for higher metallicity
    models, and that reduced strength convective overshooting is
    occuring for stars at the turnoff. At super-solar bulk
    metallicity, various indicators agree on a cluster age between
    about 3.5 and 4.0 Gyr.
\end{abstract}

\section{Introduction}

As part of the K2 M67 Study \citep{k2proj}, we are analyzing eclipsing
binary stars to provide a precise mass scale for the cluster stars as
an aid to comparisons with theoretical models. The K2 mission has uncovered
eclipsing binaries that would have been difficult to identify from the
ground, and has made it possible to precisely study even those systems
with shallow eclipses. Our goal here is to measure masses and radii to
precisions of better than 1\% in order to be comparable to results
from the best-measured binaries in the field
\citep{andersen,torres,debcat}. We have previously analyzed the
  binary WOCS 12009 \citep{s1247}, although we found that at least the
  primary star showed signs of having been part of a stellar
  merger. The brighter binary HV Cnc has recently been reanalyzed by
  \citet{gokay}, but a proper analysis is complicated by an extremely
  faint secondary star and a possibly associated third star.

WOCS 11028 (also known as Sanders 617, EPIC 211411112; $\alpha_{2000}
= 08^{\mbox{h}}50^{\mbox{m}}26\fs99$, $\delta_{2000} =
+11\degr48\arcmin31\farcs3$) in M67 was first reported as a
double-lined spectroscopic binary star by \citet{geller}, although the
system had been monitored previously for about 7 years by D. Latham
and collaborators, including determination of orbital parameters.  The
K2 M67 Study collaboration detected an eclipse in {\it Kepler} K2
observations for campaign 5. The spectroscopic parameters ($P =
62.593$ d, $e = 0.625$, $\omega = 236\degr$) showed that this meant
that the system has only one eclipse per orbit, which is not uncommon
for eccentric binaries. However, because the orbital separation is
relatively large and the eccentricity is not too extreme, the orbital
inclination can still be fairly tightly constrained between two
limits: inclinations that give two eclipses per orbit (one a grazing
eclipse), and ones that give no eclipses. For this binary, the
  singly-eclipsing range only covers inclinations from $86\fdg2$ to
  $88\fdg8$, and light-curve modeling makes it possible to constrain
  this range even further.  We will show below that we can therefore
derive precise masses for the two stars in this binary. In the
cluster's color-magnitude diagram (CMD), the binary sits near the
cluster turnoff, which is an indication that the stars are relatively
far along in their core hydrogen burning. Thus, there is hope that
these stars can significantly constrain the age of the cluster they
reside in.

In section \ref{obs}, we describe the photometric and spectroscopic
data that we collected and analyzed for the binary. In section
\ref{analy}, we describe the modeling of the binary system. In section
\ref{discuss}, we discuss the results and the interpretation of the
system.

\section{Observations and Data Reduction}\label{obs}

\subsection{K2 Photometry}

\begin{figure}
\epsscale{1.3}
% clusters/m67/s617/lcs/everest_lcs.py
\plotone{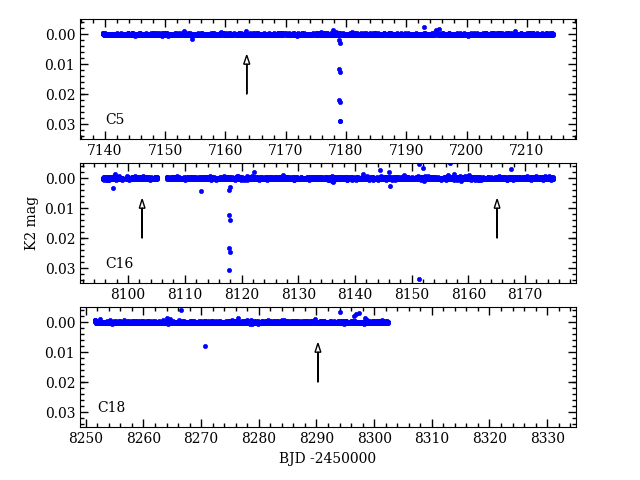}
\caption{K2 photometry from the EVEREST
  pipeline for the three campaigns that observed WOCS 11028. The two
  observed eclipses are the largest flux drops in C5 and C16,
and the predicted times of the non-eclipsing conjunction
  are shown with arrows. \label{k2phot}}
\end{figure}

M67 was observed during Campaigns 5, 16, and 18 of the K2
mission. WOCS 11028 was observed during all of the campaigns in a
custom aperture with long cadence (30 min) exposures.  Single eclipses
were observed during campaigns 5 and 16, having about a 2.7\% decrease
in flux. (No eclipse was observed in campaign 18 data, and the
ephemeris indicates that one was not expected.) The eclipse times were
measured to be at BJD $2457178.9531\pm0.00088$, and
$2458117.8789\pm0.00075$ using the bisector method of \citet{kwee}.
The radial velocity curve shows that these are secondary eclipses (of
the cooler, less massive star). In the K2 light curves, there is no
sign of primary eclipses at the four epochs predicted by the orbital
ephemeris (see Figure \ref{k2phot}).

\begin{figure}
% papers/m67/rings.v3.skycell.1534.097.stk.y.uncov.fits
\includegraphics[scale=0.45]{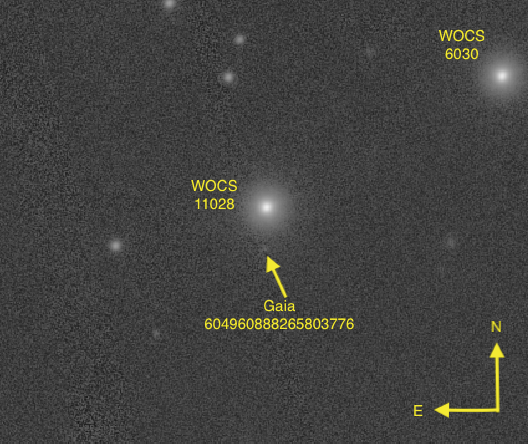}
\caption{Pan-STARRS1 $y$ image centered on WOCS 11028.  The image is
  $90\arcsec$ tall, and the sky orientation is shown.\label{yim}}
\end{figure}

Because the K2 camera has relatively large pixels ($3\farcs98$ on a
side), we briefly discuss the possibility of contaminating light from
unrelated stars. Although WOCS 11028 is a member of M67, it is in the
outskirts of the cluster. As shown in Figure \ref{yim}, the nearest
bright star is WOCS 6030 / Sanders 619 at a distance of $54\arcsec$,
and about 0.3 mag fainter in {\it Gaia} $G$ band. There is a {\it
    Gaia} source (indicated in Figure \ref{yim}) $8\arcsec$ away, but
  nearly 8 $G$ magnitudes fainter. There are three other {\it Gaia}
sources between 27 and $34\arcsec$ away, but all are at least 5.3 mag
fainter in $G$. In the light curves described below, the
  photometric apertures did not extend more than 5 pixels
  away from the photocenter, and so avoided the brighter stars. As a
result, we assume contamination of the K2 apertures used for WOCS
11028 is negligible.

Because of the incomplete gyroscopic stabilization during the K2
mission, systematic effects on the light curve are known to be substantial. We
used two different light curves for WOCS 11028 as part
of our investigation of systematic error sources.
The first light curve we experimented with utilized the K2SFF pipeline
\citep{keplerav,k2av} that involved stationary aperture photometry
along with correction for correlations between the telescope pointing
and the measured flux. The light curve that resulted still retained
small trends over the long term, so these were removed by fitting the
out-of-eclipse points with a low-order polynomial and dividing the
fit. The second light curve came from version 2.0 of the EVEREST
pipeline \citep{everest}, which uses pixel-level decorrelation to
remove instrumental effects. We used the detrended and co-trending basis
  vector-corrected light curve.
However, the eclipse depth is 
  about 4\% deeper in the EVEREST light curve than in the K2SFF
version, so we need to gauge the effects on measurements of the binary
characteristics. Because the K2 light curves have high
signal-to-noise, they have a substantial influence on best-fit binary
star parameter values (see section \ref{binary}).  The scatter among
the best-fit parameter values when comparing runs with different K2
light curve reductions is sometimes larger than the statistical
uncertainties. Because there is no clearly superior method for
processing the K2 light curves, we will take the scatter in the binary
model parameter values as a partial indicator of systematic error
resulting from the processing.

\subsection{The Spectral Energy Distribution}\label{sedsec}

Because the eclipses of the WOCS 11028 binary provide minimal
information on the sizes of the individual stars, we need other
  well-measured characteristics of the binary's stars in order to
  extract age information for the cluster. Fortunately, M67 has been
heavily observed over a wide wavelength range, making it possible to
construct well-sampled spectral energy distributions for the binary
and similarly bright stars in the cluster. In this section, we describe the
spectroscopy and photometry we have assembled, and the efforts to put
the observations on a consistent flux scale.

{\it Ultraviolet:} We obtained photometry and a spectrum from the {\it
  Galaxy Evolution Explorer} ({\it GALEX}; \citealt{galex}) archive
for the NUV passband ($1771-2831$ \AA). WOCS 11028 was imaged twice,
for 1691.05 s (GI1 proposal 94, P.I. W. Landsman) and 5555.2 s (GI1
proposal 55. P.I. K. Honeycutt).
As the archived magnitudes are based
on count rates with minimal background contributions, we computed a
final magnitude and flux based on the average count rate for both
observations.  \citet{galexcal} describes the characteristics of the
GALEX photometry and its calibration to flux. GALEX magnitudes are on
an AB system \citep{oke}, and we used the zero point magnitude
($m_{NUV} = 20.08$) and reference flux ($2.06 \times 10^{-16}$ erg
s$^{-1}$ cm$^2$ \AA$^{-1}$) to convert to flux.

A NUV grism spectrum was taken as part of GI1 proposal 94 (PI:
W. Landsman) on 2005 Feb. 18 with a total exposure time of 27653 s.
The NUV spectrum covers a wider wavelength range than the NUV
photometry filter, but there was effectively no signal detected for
WOCS 11028 at wavelengths less than about 2000 \AA. The spectrum was
obtained from MAST, and the flux calibration from the pipeline
reduction was used.

Observations were taken of a smaller portion of M67 (including WOCS
11028) using the UVOT telescope on the Swift satellite. We collected
UV fluxes in the $uvw1$, $uvm2$, and $uvw2$ bands from the Swift UVOT
Serendipitous Source Catalogue (version 1.1; \citealt{swiftssc}).
The three bands mostly cover the same wavelength range as the GALEX
NUV filter, but with somewhat finer resolution. Flux correction
factors from \citet{uvotcal} were applied to go from the gamma-ray
burst spectrum calibration given in the archive to one utilizing
\citet{pickles} library stars.

{\it Near-ultraviolet, Optical and Near-infrared:} M67 has been frequently observed from
the ground for the purposes of photometric calibration. For the
purposes of an SED, narrow-band filters are particularly useful, and
we discuss these first. \citet{balaguer} presented Str\"{o}mgren
$uvby$ photometry for the cluster. Because M67 stars are commonly used
as standards in Str\"{o}mgren photometry \citep{nissen}, we can be
fairly assured that the observations are tied to the standard system.
We employed reference fluxes from \citet{gray-strom} to convert the
magnitudes to fluxes.

\citet{fan} conducted wide-field observations of M67 in a series of
narrow-band filters (``BATC'') covering from 3890 to 9745 \AA ~ using
bandpasses avoiding most important sky lines. The wavelength coverage
of the filters is better in the near-infrared, making them a good
complement to Str\"{o}mgren photometry. The BATC survey goal was
spectrophotometry at the 1\% level, and the study largely achieved
that, judging from the low scatter in their color-magnitude diagrams.
Their reported magnitudes are on the \citeauthor{oke} AB system. We
found that fluxes from these magnitudes were systematically higher
than similar observations in other systems. We collected magnitudes in
a larger set of filters from BATC Data Release
1\footnote{http://vizier.u-strasbg.fr/viz-bin/VizieR-3?-source=II/262/batc},
and found that these were more consistent, probably due to improved calibration
\citep{zhoubatc}.

Narrow-band photometry is also available in the Vilnius filter system
for many stars in M67 (although not WOCS 11028) in the study of
\citet{laug}. We converted the reported magnitudes to fluxes using the
zeropoints from \citet{mvb}. The filters in the Vilnius system are
somewhat denser in the blue portions of the optical, and these again
complement the spectrophotometry in the BATC system.

We have Johnson-Cousins photometry in $BVI_C$ from \citet{mephot} and
\citet{yadav}, in $UBVR_CI_C$ from \citet{mmj}, in $BVRI$ from
\citet{nardiellom67}, and in $BV$ from Data Release 9 of the AAVSO
Photometric All-Sky Survey (APASS; \citealt{apass}). These
measurements are on a Vega magnitude system, and so have been
converted to fluxes using reference magnitudes from Table A2 of
\citet{bcp}, and accounting for the known reversal of the zero point
correction rows for $f_\lambda$ and $f_\nu$. The absolute calibrations
of each of these studies are unavoidably different for the same filter
bands, and this will contribute to the noise in the SEDs. However,
this does not affect their use in determining the relative
contributions of the two stars in the WOCS 11028 binary (see \S
\ref{sedspec}). (The $R$ and $I$ filter observations given in
  \citet{nardiellom67} were actually taken in SDSS $r$ and $i$
  filters, and have been recalibrated to the Sloan DR12 system
  (D. Nardiello, private communication).  These data were not used in
  fits to SEDs.)

There are several additional surveys that provide calibrated broad-band
photometric observations. We used PSF magnitudes from Data Release 14
of the Sloan Digital Sky Survey (SDSS; \citealt{sdss}), calibrated
according to \citet{sdsscal}. The SDSS is nearly on the AB system,
with small offsets in $u_{SDSS}$ and $z_{SDSS}$ that we have also corrected for
here.  The Pan-STARRS1 survey \citep{ps1desc} contains photometry in 5
filters, and we use their mean PSF magnitudes here. Zero points for
its AB magnitude system are given in \citet{ps1}. The APASS survey
also observed the cluster in Sloan $g^{\prime} r^{\prime} i^{\prime}$
filters, and their photometry was flux calibrated from the AB system.

Finally, {\it Gaia} has already produced high-precision photometry
extending far down the main sequence of M67 as part of Data Release
2. We obtained the fluxes in the $G$, $G_{BP}$, and $G_{RP}$ bands
from the {\it Gaia} Archive.

{\it Infrared:} We obtained Two-Micron All-Sky Survey (2MASS;
\citealt{2mass}) photometry in $JHK_s$ from the All-Sky Point Source
Catalog, and have converted these to fluxes using reference fluxes for
zero magnitude from \citet{2masscal}. Many stars also have photometry in
three bands from the Wide Field Infrared Explorer (WISE;
\citealt{wise}), which were also converted to fluxes using tabulated reference
fluxes at zero magnitude.

\subsubsection{Photometric Deconvolution}\label{sedspec}

\begin{figure*}
\epsscale{1.3}
% clusters/m67/sss/plotcmds.py
\plotone{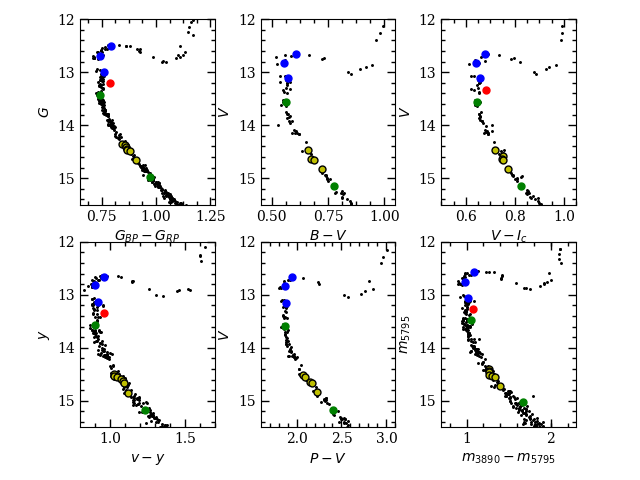}
\caption{Color-magnitude diagrams of likely single stars near the
  turnoff of M67 in photometric datasets with the highest
  signal-to-noise. ($BV$ and $VI_c$: \citealt{mephot}; $vy$:
  \citealt{balaguer}; $PV$: \citealt{laug}; $m_{3890}m_{5795}$:
\citealt{fan}).
  The red points are combined photometry of the
  WOCS 11028 binary, green points are the proxy photometry values for
  the components of the binary, yellow points are solar twins, and
  blue points are stars identified as being at critical points in the
  evolutionary sequence.
  \label{turnoffs}}
\end{figure*}

A benefit of the binary's membership in the M67 cluster is that it should
be possible to describe the binary's light as the sum of the light of
two single cluster stars. To that end, we compiled a database of
photometric measurements from likely single main sequence stars in
M67, and sought a combination of stars whose summed fluxes most closely match
the fluxes of WOCS 11028. For our sample of probable single stars, we
selected likely members based on {\it Gaia} proper motions,
parallaxes, and photometry, as well as radial velocity membership and
binarity information from \citet{geller}. Likely binaries were
rejected by restricting the sample to those classified as radial
velocity SM (single members) with {\it Gaia} photometry placing them
within about 0.03 mag of the blue edge of the main sequence band in
the $G_{BP}-G_{RP}$ color. Selected stars are shown in Figure \ref{turnoffs}.

The benefit of this procedure for constraining the SEDs of the
  binary's stars is that it is a {\it relative} comparison using
other cluster stars with the same distance, age, and chemical
composition. As such, it is independent of distance and reddening (as
long as these are the same for the binary and comparison stars), the
details of the filter transmission curves (as long as the same filter
is used for observations of the different M67 stars), and flux
calibration of any of the filters (as long as the calibration is
applied consistently). We can also avoid systematic errors associated
with theoretical models or with the consistency of the different parts
of empirical SEDs compiled from spectra.

We tested two ways of doing the decomposition: using actual M67 stars
as proxies and checking all combinations of likely main sequence
stars; and fitting all main sequence stars with photometry in a given
filter as a function of {\it Gaia} $G$ magnitude in order to derive SEDs
that could be combined. When using actual M67 stars, we are somewhat
at the mercy of the photometry that is available for each star (and
the binary) and of the stellar sampling of the main sequence. The use
of fits allows for finer examination of the main sequence, although
there is some risk of diverging from the photometry of real stars.
Even for a relatively rich cluster like M67, parts of the main
sequence are not well populated, and we believe that the main-sequence fitting
method gives better results in that case. We present the results of
both analyses below, however.

To judge the agreement between the summed fluxes of a pair of stars and the binary
photometry, we looked for a minimum of a $\chi^2$-like parameter
involving fractional flux differences in the different filter bands:
\[ \sum_i \left(\frac{F_{i,bin} - (F_{i,1} + F_{i,2})}{\sigma_{i,bin} \cdot F_{i,bin}}\right)^2 \]
where $\sigma_{i,bin} = 10^{-(\sigma_{i,m}+0.01)/2.5} - 1$ and
$\sigma_{i,m}$ is the magnitude uncertainty in the $i$th filter band
for the binary.  The addition of 0.01
mag somewhat deweights photometry with very low uncertainties ({\it
  Gaia} and GALEX NUV) that results partly from their very wide filter
bandpasses.

\begin{figure}
\epsscale{1.3}
  % clusters/m67/sss/sedcompfig.py
\plotone{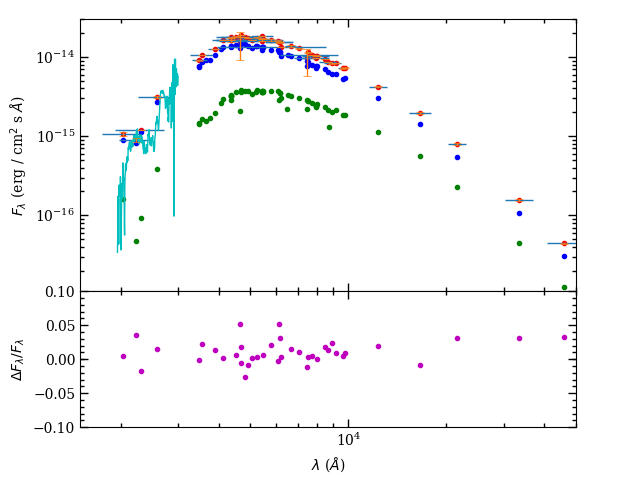}
\caption{{\it Top:} SED for the binary WOCS 11028 (photometry: red
  points, GALEX spectrum: cyan lines) and for the main sequence stars
WOCS 6018 (blue points) and WOCS 16013 (green points). Horizontal error bars represent the effective
width of the filter. {\it Bottom:} Fractional difference between the
binary star photometry fluxes and the result of the main sequence fitting procedure.\label{sedfig}}
\end{figure}

When using M67 stars as proxies, the results can be affected by the
selection of filters that could be used. We examined solutions excluding the {\it
  Swift}/UVOT and/or Str\"{o}mgren photometry because they covered the
smallest portion of the cluster field, and excluding them allowed us
to use larger sets of stars. The sample sizes were 52 stars
having photometry in all of the filter bands, 109 with UVOT excluded,
and 125 with UVOT and Str\"{o}mgren photometry excluded.  In all
cases, the best fits involved WOCS 6018 (Sanders 763) as the brighter
star, while the preferred fainter star was either WOCS 10027/S1597
(with UVOT photometry excluded) or WOCS 16013/S795 (for all photometry,
or with UVOT and Str\"{o}mgren excluded). Solutions involving WOCS
6018 with different faint stars (WOCS 17021/S820, WOCS 21046/S1731,
WOCS 16018/S814) were the next-best fits, and all of these stars reside
in similar positions on the main sequence.
Figure \ref{sedfig} shows a comparison of the SED of WOCS 16013 with
WOCS 6018, giving an indication of the significant optical and
infrared excess for the binary. The bottom panel shows the residuals
between the binary flux and the summed fluxes of WOCS 6018 and WOCS
10027. A potential limiting factor is the stellar
sampling available near the brighter star, but we have stars within
0.007 $G$ mag on the bright side and within 0.016 mag on the faint
side.

\begin{figure}
\epsscale{1.3}
  % clusters/m67/sss/sumfits.v2.py
\plotone{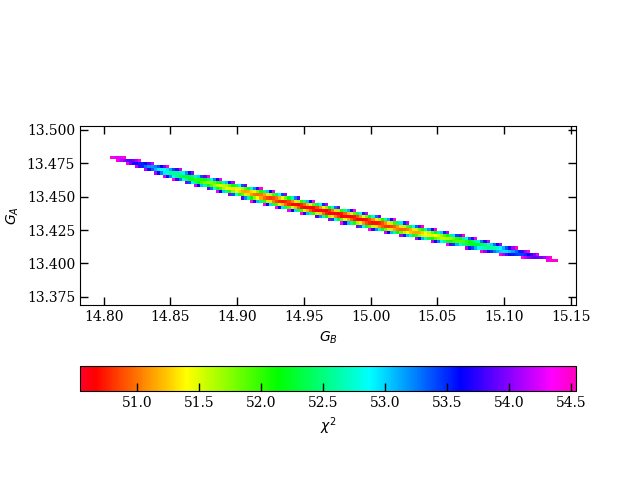}
\caption{Goodness-of-fit contours in goodness-of-fit statistic $\chi^2$ versus {\it Gaia} $G$ magnitudes for the
  two components of WOCS 11028 out to approximately $2\sigma$ away
  from the best fit.
\label{sedcontour}}
\end{figure}

The main-sequence fitting procedure can be employed in any filter with
a sufficient sample of stars covering the range of brightnesses for
the binary's stars. In our case, this eliminated the $R_C$, {\it Swift}
$uvm2$, and WISE W3 and W4 filters from consideration.
Our fit statistic had a minimum value of 50.0 for the selection of 46
filters.  We estimated the $2\sigma$ uncertainty in the fit based on
where the goodness-of-fit statistic reached a value of 4 above the
minimum value. For example, this returns $2\sigma(G_A) = 0.035$ and
$2\sigma(G_B) = 0.15$.  As expected, there is an anti-correlation
between values for the primary and secondary stars because of the need
to match the binary fluxes (see Fig. \ref{sedcontour}).

\begin{figure}
\epsscale{1.3}
  % clusters/m67/sss/galexspecfig.py
\plotone{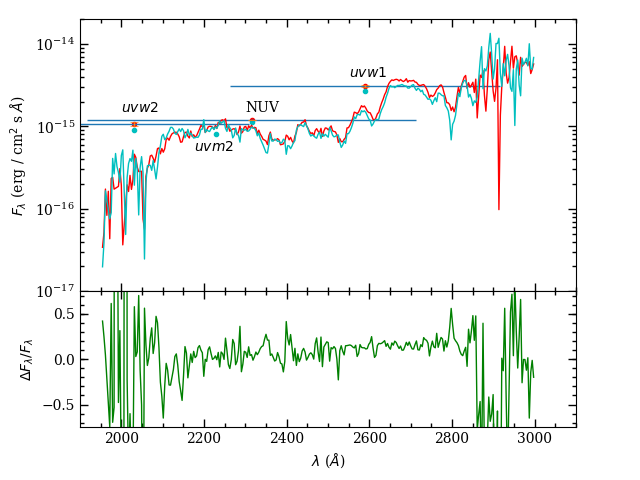}
\caption{{\it Top:} Comparisons of GALEX NUV spectra for the binary
  WOCS 11028 ({\it red}) and likely main sequence star WOCS 6018 ({\it
    cyan}). Photometry from GALEX (NUV filter) and {\it Swift}/UVOT is
  shown as points, with the approximate widths of the filters shown as
  horizontal bars. {\it Bottom:} Fractional difference between GALEX
  spectra of WOCS 11028 and WOCS 6018.\label{galspec}}
\end{figure}

For an additional check on the brighter star, we compared the GALEX
ultraviolet spectrum of WOCS 11028 with that of WOCS 6018, as shown in
Fig. \ref{galspec}.  Although our minimization procedure encourages
agreement in near-UV filters, it doesn't require agreement of the
spectra. In spite of this, the overall shape and flux level of the two
stars agree well, with the binary becoming consistently higher on the
long wavelength end. The slope of the GALEX spectrum is one of the
more sensitive indicators of temperature for stars on the upper main
sequence of M67, and this combination of stars does a better job of
reproducing that than two equal-mass stars, for example.  The
enhancement in flux at the long-wavelength end of the spectrum can be
attributed to a small contribution from the faint star in the
binary. The four photometric observations in the near ultraviolet (from
GALEX and {\it Swift}/UVOT) indicate that the binary is slightly
brighter than WOCS 6018 in all of the filter bands.

The above fits to the binary's photometry provide luminosity
ratios in different bands independent of models, and we use some of
these as constraints in modeling the radial velocity and eclipse light
curve data in \S \ref{binary}. These ratios are provided in Table
  \ref{sedtable}.

\subsubsection{Effective Temperatures and Bolometric Fluxes}

\begin{figure}
\epsscale{1.3}
% clusters/m67/sss/fitphot.py
\plotone{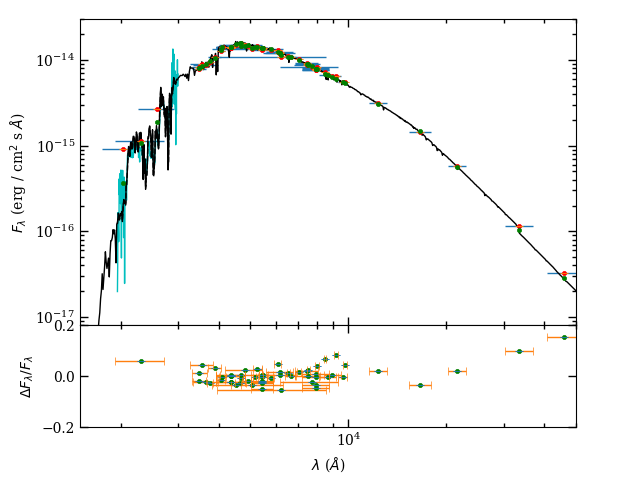}
\caption{{\it Top:} SED for the proxy for the bright star in the WOCS
  11028 binary (red points for photometry) and a fitted ATLAS9 model
  for $T_{eff} = 6185$ K and $\log g = 4.25$ (solid line, and green
  points for integrations over filter response curves). Horizontal
  error bars represent the effective width of the filter. A GALEX
  spectrum of WOCS 6018 is shown in cyan. {\it Bottom:} Fractional
  difference between the stellar fluxes and the best-fit model fluxes
  for ATLAS9 (green)
  \label{brightfig}}
\end{figure}

\begin{figure}
\epsscale{1.3}
% clusters/m67/sss/fitphot.py
\plotone{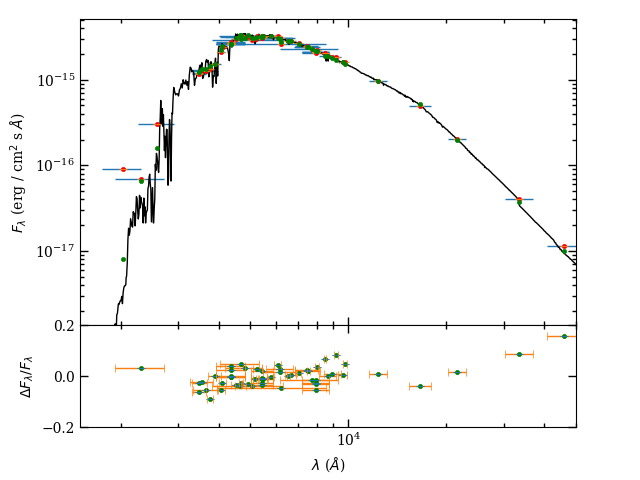}
\caption{{\it Top:} SED for the proxy for the faint star in the WOCS 11028 binary (red
  points) and a fitted ATLAS9 model for $T_{eff} = 5500$ K (solid
  line, and green points for integrations over filter response
  curves). Horizontal error bars represent the effective width of the
  filter. {\it Bottom:} Fractional difference between the stellar
  fluxes and the best-fit model fluxes for ATLAS9 (green)
  models.\label{faintfig}}
\end{figure}

With SEDs in hand for stars that we believe are good proxies for the
binary's components, we can derive additional properties via fits with
theoretical models. The models can introduce systematic errors in
  the quantities we try to measure, although we will try to mitigate
  them. We tested models from ATLAS9 \citep{atlas9}\footnote{The
  models were calculated using the ATLAS9 fortran code that employed
  updated 2015 linelists, and at temperatures between the published
  gridpoints.} and \citet{coelho14}, but the results were very
similar, and we primarily discuss ATLAS9 model fits below. Models were
adjusted to account for the interstellar reddening of M67
[$E(B-V)=0.041\pm0.004$; \citealt{taylor}] using the \citet{cardelli}
extinction curve. Model photometry was calculated using the
IRAF\footnote{IRAF is distributed by the National Optical Astronomy
  Observatory, which is operated by the Association of Universities
  for Research in Astronomy (AURA) under a cooperative agreement with
  the National Science Foundation.} routine {\tt sbands}. Bolometric
flux was derived from the median multiplicative factor needed to bring
the best-fit model photometry into agreement with each observation,
and we quote an uncertainty on the median that is based on the range
covered by observations that are within $\sqrt{N}/2$ entries of the
middle in the ordered list.
We find that the SEDs of WOCS 6018 and the bright component
from our main sequence fit method are in good agreement with models of
6200 K
(see Fig. \ref{brightfig}),
while all of the candidate fainter
components indicate a temperature of about 5500 K (see
Fig. \ref{faintfig}).

To try to achieve greater precision, we also calculated temperatures
using the infrared flux method \citep[][IRFM;]{irfm}. Briefly, this
method exploits the difference in temperature sensitivity between the
bolometric flux and monochromatic fluxes in the infrared on the
Rayleigh-Jeans portion of the spectrum. With the available photometry
databases, we have measurements of fluxes covering the majority of the
stellar energy emission. The ratio of the bolometric and infrared
fluxes (we use fluxes in 2MASS bands) can be compared to theoretical
values (where we again use ATLAS9 models):
\[ \frac{\mathcal{F}_{\rm bol}(\mbox{Earth})}{\mathcal{F}_{\lambda_{IR}}(\mbox{Earth})} = \frac{\sigma T_{\rm eff}^4}{\mathcal{F}_{\lambda_{IR}}(\mbox{model})} \]
We used the 2MASS flux calibration of \citet{casairfm} in this case,
in part because it produced greater consistency between the
temperatures derived in the three bands --- full ranges between 15 and
40 K. Starting from a solar-metallicity ATLAS9 model that produced a
good fit by eye, we adjusted the temperature of the synthetic spectrum
until it matched the average IRFM temperature from the three 2MASS
bands. We find temperatures of 6185 K and 5500 K for the two
stars. The model surface gravity was chosen from the eclipsing binary
results or from MESA models, although changes had little effect.
Uncertainty in the reddening (which modifies the shape of the
theoretical model we fit to the data) and metal content of the models
affects the measured temperature at about the 15 K level. Overall, we
estimate that there is an uncertainty of about 50 K and 75 K in the
temperatures.

For comparison, there are spectroscopic temperatures
available for stars near the position of the brighter star in the CMD
from recent surveys looking at abundance differences as a function of
evolutionary phase.  \citet{liu} found temperatures by forcing
excitation and ionization balance in their modeling of Fe I and II
lines, and for three stars slightly brighter in $G$, they derived
temperatures between about 6100 and 6150 K.  \citet{souto19} found
temperatures from ASPCAP fits to infrared APOGEE spectra and the three
closest stars in $G$ covered a range from 6050 to 6110 K (in their
``calibrated'' ASPCAP values).  \citet{gao} derived temperatures from
fits to first and second ionization states of Sc, Ti, and Fe lines,
and two H Balmer lines, and for the stars nearest in $G$ magnitude,
they found a range from about 6090 to 6140 K. (Our best fit M67 star,
WOCS 6018, had a temperature of 6127 K.) \citet{b_m} used temperatures
from the {\it Gaia}-ESO Survey, and observed four stars slightly
fainter than WOCS 6018 that returned a range between 6000 and 6060
K. \citet{onehag} derived temperatures from photometry, but also
examined ionization and excitation temperatures. For five stars
slightly fainter than our fit, they found temperatures between about
6130 and 6200 K, with the closest matches in brightness being at the
high end of that range.  The agreement of our IRFM temperature with
these spectroscopic measurements is quite good, and if anything, our
temperature is higher than spectroscopic values by $\sim 50-100$ K.

We can also derive bolometric fluxes at Earth from the model fits to
the SEDs. The SEDs for both stars are very well sampled
with photometry, and a GALEX spectrum for the best
bright star proxy can be employed as well.
The results are summarized in Table \ref{sedtable}.
The SED model fits can be combined to
  provide a constraint on the radius ratio for the two stars that is
  independent of the distance and reddening of the cluster via
\[ \frac{R_2}{R_1} = \sqrt{\frac{F_{bol,2}}{F_{bol,1}} \cdot \left(\frac{T_1}{T_2}\right)^4} \]
Because the single grazing eclipse per orbit gives us a constraint on
the sum of the stellar radii, this radius ratio will allow us to
compute the individual stellar radii.

\subsection{Spectroscopy}\label{spectro}

We have measured radial velocities for the WOCS 11028 binary using
three spectroscopic datasets.  The first and largest set we employ
here comes from the CfA Digital Speedometers
\citep{latham85,latham92} on the 1.5 m Tilinghast reflector at Fred
Lawrence Whipple Observatory and the MMT. These observations were
recordings of a single echelle order covering 516.7 to 521 nm around
the Mg I b triplet using an intensified photon-counting Reticon
detector.
The spectra were taken as part of a larger monitoring campaign of M67
stars.

We made two observations of the binary using the HARPS-N spectrograph
\citep{harpsn} on the 3.6 m Telescopio Nazionale Galileo
(TNG). HARPS-N is a fiber-fed echelle that has a spectral resolving power
$R = 115000$, covering wavelengths from 383 to 693 nm. The spectra were
processed, extracted, and calibrated using the Data Reduction Software
(version 3.7) provided with the instrument.

We also obtained five archival spectra from the APOGEE database
\citep{apogee} that were taken between January and April 2014.  These
are $H$-band infrared ($1.51 - 1.70 \mu$m) spectra with a spectral
resolution $R \sim 22500$ taken on the 2.5 m Sloan Foundation
Telescope.  We used APOGEE flags to mask out portions of the spectrum
that were strongly affected by sky features, and continuum normalized
the spectra using a median filter.

The radial velocities were measured using the spectral separation
algorithm described in \citet{gonz}.  In the first iteration step, a
master spectrum for each component is isolated by aligning the
observed spectra using trial radial velocities for that component and
then averaging. This immediately de-emphasizes the lines of the
non-aligned stellar component, and after the first determination of
the average spectrum for each star, the contribution of the
non-aligned star is subtracted before the averaging in order to better
clean each spectrum. The radial velocities can also be remeasured from
spectra with one component subtracted. This procedure is repeated for
both components and continued until a convergence criterion is met.
We measure the broadening functions \citep{rucinski02} to determine
the radial velocities using narrow-lined synthetic spectra as
templates. Radial velocity measurements from broadening functions
improve accuracy in cases when the lines from the two stars are
moderately blended \citep{rucinski02}.

The CfA spectra covered a wide range of orbital phases that made it
possible to use a large number of spectra in calculating average spectra for the
two stars. In most cases, spectra were left out due to low
signal-to-noise ratio. We used 17 out of 28 spectra in
determining the average spectrum for the primary, but only used the 13
spectra with the most clearly detected secondary component to
determine its average spectrum. With good average spectra, it was
possible to subtract each component out of observed spectra even when
they were taken close to crossing points (phases $\phi \approx 0.05$
and 0.6). After some experimentation, we found that
synthetic templates (from the grid of \citet{coelho05}) of 5250 K
optimized the detection of broadening function peaks for the secondary
star. A template with 6250 K was used for the primary star. For the
CfA velocities only, we used run-by-run corrections derived from
observations of velocity standards.  Run-by-run offsets measured for
the CfA spectra ranged from $-1.52$ to $+2.21$ km s$^{-1}$. We
initially assigned velocity uncertainties derived from the
spectral separation analysis, but scaled these to ensure that the scatter
around a best-fit model for this dataset was consistent with the
measurement errors. (In other words, we forced the reduced $\chi^2$
value to 1.) The average uncertainty for the primary star velocities
was 1.25 \kms, and for the secondary star it was 4.41 km s$^{-1}$.

The APOGEE spectra were analyzed separately, and detection of the
secondary star features were considerably more secure. The synthetic
spectral templates were taken from ATLAS9/ASS$\epsilon$T models
\citep{asset} calculated for the APOGEE project.  The uncertainty
estimate for each APOGEE velocity was generated from the rms of the
velocities derived from the three wavelength bands in the APOGEE
spectra. This was typically around 0.15 \kms for the primary star and
0.65 \kms for the secondary star. The velocity was corrected
to the barycentric system using values calculated by the APOGEE
pipeline. For all five spectra, the broadening function peaks for the
two stars were resolvable thanks to the high resolution of the spectra
and fortuitous phases of observation.

Because we only had two observations using the HARPS-N spectrograph
and one of them was taken at a phase very near a crossing point, we
determined velocities using broadening functions alone. To get an
internal measure of the velocity uncertainties, we measured the velocities in
four 30 nm subsections of the spectra and computed the error in the mean. 

Because the synthetic spectra used in our broadening function
measurements do not account for gravitational redshifts of the stars,
our velocities should have an offset relative to velocities derived
using an observed solar template. For the CfA spectra, the
  run-by-run corrections were computed using dawn and dusk sky
  exposures, and this defines the native CfA velocity system used in
  \citet{geller}, incorporating the gravitational redshift of the Sun.
  For consistency, we subtracted the combined gravitational redshift
  of the Sun and gravitational blueshift due to the Earth (0.62 km
  s$^{-1}$) from the APOGEE and HARPS-N velocities we
  tabulate. Based on the derived masses and radii for the stars in the
  binary, they should have slightly larger gravitational redshifts
  than the Sun (0.66 and 0.69 km s$^{-1}$, versus 0.64 \kms for the
  Sun) but we did not correct for these smaller differences. In our
  later model fits, we did, however, allow the system velocities for
  the two stars to differ to account for effects like this and
  convective blueshifting that could produce small systematic
  shifts. These could affect the measured radial velocity amplitudes
  $K_1$ and $K_2$ (and the stellar masses) if not modeled. 

\begin{figure}
\epsscale{1.3}
% clusters/m67/s617/elc/rvsELC_sep.py
\plotone{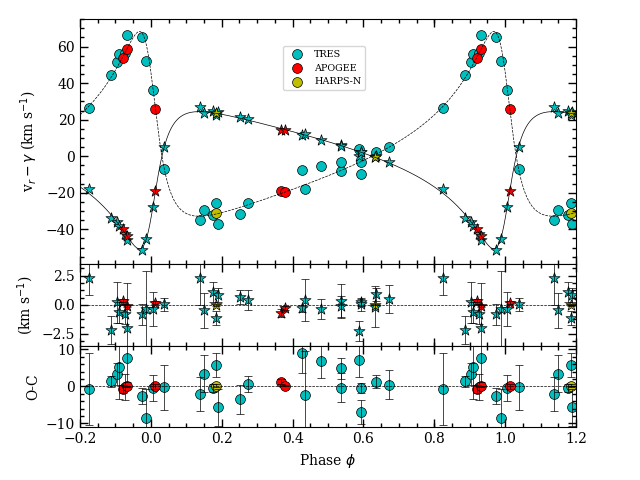}
\caption{Phased radial velocities (minus system velocity) for WOCS 11028, along with the
  best-fit model. Measurements for the primary and secondary stars are
  shown with stars and circles, respectively.  The lower panels show
  the observed minus computed values with error bars scaled to give a
  reduced $\chi^2=1$ (see \S \ref{binary}).\label{rvplot}}
\end{figure}

 The measured radial velocities are presented in Table \ref{spectab},
 and the phased radial velocity measurements are plotted in
 Fig. \ref{rvplot}. The model fits are described in Section 3.4,
   but the parameters that were used to fit the velocity data were the
   velocity semi-amplitude of the primary star $K_1$, mass ratio
   $q=M_2 / M_1=K_1/K_2$, eccentricity ($e$), argument of periastron
   ($\omega$), and systematic radial velocities $\gamma_1$ and
   $\gamma_2$. In the end, we do not see noticeable
 differences between the systematic velocities.

% s617.rv(ab).disent.scalecorr.bjd.dat
\begin{deluxetable*}{rcccc|rcccc}
\tablewidth{0pt}
\tabletypesize{\scriptsize}
\tablecaption{Radial Velocity Measurements}
\tablehead{\colhead{mJD\tablenotemark{a}} & \colhead{$v_A$} & \colhead{$\sigma_{A}$} & \colhead{$v_B$} & \colhead{$\sigma_B$} & \colhead{mJD\tablenotemark{a}} & \colhead{$v_A$} & \colhead{$\sigma_{A}$} & \colhead{$v_B$} & \colhead{$\sigma_B$}\\
& \multicolumn{2}{c}{(km s$^{-1}$)} &  \multicolumn{2}{c|}{(km s$^{-1}$)} & &  \multicolumn{2}{c}{(km s$^{-1}$)} &  \multicolumn{2}{c}{(km s$^{-1}$)}}
\startdata
\multicolumn{5}{c|}{CfA Observations}       & 49678.02714 & 54.71 &  0.59  &  6.53 & 3.23\\
48347.73537 & $-13.31$ &3.84 & 98.95 & 6.47 & 49700.00054 & 38.07 &  0.89  & 29.41 & 2.87\\
48618.00088 & 53.87 &  0.59  &  0.83 & 3.95 & 49705.98524 & 31.68 &  1.77  & & \\
48635.87968 & 38.55 &  1.48  & 24.19 & 3.95 & 50535.69747 & $-1.66$ &1.18  & 76.62 & 1.44\\
48676.79118 & 56.54 &  0.59  &$-4.62$& 5.03 & 50536.76047 & $-3.37$ &1.77  & 84.15 & 2.87\\
48701.71198 & 32.68 &  0.89  & 36.16 & 4.67 & 50885.78158 & 41.13 &  0.89  & 27.03 & 4.67\\
48704.66488 & 32.50 &  0.59  & 34.70 & 1.80 & 50913.75637 & $-10.46$ &0.89 & 88.80 & 3.59\\
48726.74348 & $-13.03$ &3.25 & 84.47 & 7.90 & 50916.72187 & $-18.69$ &0.89 & 97.47 & 2.16\\
48754.67588 & 44.44 &  1.77  & 14.28 &11.86 & 50918.72237 &  4.69 &  1.48  & 68.75 & 3.59\\
48942.01949 & 44.13 &  0.30  & 25.07 & 5.39 & 50920.71047 & 37.55 &  0.59  & 25.23 & 6.11\\
48966.93949 & 14.32 &  1.48  & 58.96 & 9.70 & \multicolumn{5}{c}{APOGEE Observations}\\
48988.95240 & 57.20 &  0.89  &$-0.02$& 1.08 & 56672.86368 & $-11.57$ & 0.06 & 91.22 & 0.09 \\ % APOGEE (disentangling, gr correction)
49019.89560 & 29.34 &  1.18  & 37.60 & 3.95 & 56677.88083 &  13.45 & 0.15 & 57.87 & 0.64\\
49049.81820 & 56.23 &  1.48  &  3.07 & 5.03 & 56700.76762 &  46.59 & 0.15 & 12.91 & 0.17\\
49057.71970 & 52.50 &  0.89  &  6.60 & 2.16 & 56734.66627 & $-7.61$ & 0.13 & 85.78 & 1.07\\
49077.68900 & 34.51 &  0.59  & 29.15 & 1.44 & 56762.63514 &  46.73 & 0.35 & 13.13 & 0.64\\
49111.72870 & 59.12 &  2.36  &$-2.48$& 4.67 & \multicolumn{5}{c}{HARPS-N Observations}\\    
49328.01792 & 34.75 &  0.30  & 22.47 & 3.23 & 57752.76085 & 55.83 & 0.08 & 1.04 & 0.30 \\% HARPS, BF only, gr correction
49347.86222 & $-5.88$ &0.89  & 88.10 & 8.62 & 57780.74729 & 31.82 & 0.03 & & \\
\enddata
\label{spectab}
\tablenotetext{a}{mJD = BJD - 2400000.}  
\end{deluxetable*}

It is always prudent to check for systematic differences in velocity
when utilizing datasets from more than one observational set-up. Among
the measured primary star velocities, mean residuals (observed minus
computed) were $-0.14$ \kms for APOGEE, $+0.02$ \kms for CfA, and
$0.02$ \kms for HARPS-N. For the secondary star, we found mean
residuals of $+0.78$ \kms for CfA, $+0.10$ \kms for APOGEE, and
$+0.09$ \kms for HARPS-N (one measurement). Of these, only the average
residual for the CfA secondary velocities was more than one standard
devation (0.20 km s$^{-1}$) away from zero.
Because the APOGEE and HARPS-N spectra have higher
  signal-to-noise than the CfA spectra, they are the main contributors
  to measurement of the radial velocity amplitude of the secondary
  star, which critically affects the calculated mass of the primary star.

\section{Analysis}\label{analy}

\subsection{Cluster Membership}

WOCS 11028 is a fairly large distance (about $13\farcm7$) from
  the center of M67, but \citet{carrera} find that half of the cluster
  turnoff stars fall within $12\farcm5$ of the cluster center. As a
  result, its position is not strong evidence of being a field star.

All ground-based proper motion studies
\citep{sanderspm,girardpm,zhaopm,yadav,nardiello} indicate a high
probability ($\ge 90$\%) of cluster membership for WOCS 11028, with
the exception of \citet{kkpm}, who give 10\%). {\it Gaia} observations from Data
Release 2 \citep{GaiaDR2} have produced much more precise proper
motions recently, and the binary still has a proper motion vector
($\mu_\alpha = -11.08\pm0.04$ mas yr$^{-1}$, $\mu_\delta =
-3.14\pm0.04$ mas yr$^{-1}$) safely residing among other likely
members (centered at $\mu_\alpha = -10.97$ mas yr$^{-1}$, $\mu_\delta
= -2.94$ mas yr$^{-1}$; \citealt{GaiaHR}). Similarly, the {\it Gaia}
parallax ($\omega = 1.111\pm0.024$ mas) is within $1\sigma$ of the cluster mean
($\bar{\omega} = 1.132$ mas).

\citet{geller} published membership probabilities based on their
radial velocity survey by comparing the velocity distributions of
cluster members and field stars. They classified WOCS 11028 as a
binary member (98\% probability) based on their system velocity
($32.91\pm0.17$ km s$^{-1}$) measured from the CfA observations and
carefully placed on the zeropoint of other stars in the field.  The
best-fit system velocity from our measurements ($\gamma = 32.37$ km
s$^{-1}$) is slightly off from the mean cluster velocity of
\citeauthor{geller} (33.64 \kms with a radial velocity dispersion of
$0.59^{+0.07}_{-0.06}$ km s$^{-1}$). However, even with our lower
$\gamma$ velocity, the radial velocity membership probability is still
95\%.

Based on the three-dimensional kinematic evidence, the binary is a high
probability cluster member. 

\subsection{Distance Modulus}\label{dm}

With the release of {\it Gaia} DR2 parallaxes, the M67 distance
modulus should be revisited, as it will play a large role in
comparisons with models later in the article.  The mean parallax
derived for cluster members from {\it Gaia} DR2 measurements
($\bar{\omega} = 1.1325\pm0.0011$ mas; \citealt{GaiaHR}) is very
precise, but possibly affected by a zeropoint
uncertainty. \citet{lindegren} finds an offset of about $-30 \: \mu$as
(with tabulated parallaxes being smaller than reality), while
systematic variations with position, magnitude, and color were below
$10 \: \mu$as. At the other extreme, \citet{st18} found an offset of 
$-82\pm33 \: \mu$as using a sample of eclipsing binaries.
\citet{zinn} find offsets of $-52.8$ and $-50.2 \:
\mu$as using comparisons with asteroseismic giants and red clump stars
in the {\it Kepler} field.  \citet{schonrich} derived an offset of
$-54\pm6 \:\mu$as from an analysis of all stars (more than 7 million)
in the radial velocity sample of {\it Gaia} DR2.

These indications 
are consistent with the situation for M67. Without correction, the {\it
  Gaia} cluster mean distance modulus would be $(m-M)_0 = 9.730\pm0.002$, but this
is comparable to what has been derived previously for the distance
modulus {\it with} extinction. For example, \citet{pasquini} derive
$(m-M)_V = 9.76\pm0.06\pm0.05$ (statistical and systematic
uncertainties) from ten solar analogs, \citet{mephot} derived
$9.72\pm0.05$ using metal-rich field dwarfs with {\it Hipparcos}
parallaxes, and \citet{stellom67k2} found $9.70\pm0.04$ from K2
asteroseismology of more than 30 cluster red giants. With a $V$-band
extinction $A_V \approx 0.12$ expected, this is a significant
discrepancy. For subsequent calculations in this article, we will
assume a $-54 \: \mu$as offset in the Gaia parallaxes and include a
systematic uncertainty of $30 \: \mu$as to conservatively account for
disagreement in the parallax offsets in the literature. The resulting
distance modulus is $(m-M)_0 = 9.63\pm0.06$.

\subsection{Li Abundance and Stellar Rotation}\label{liline}

Even though the WOCS 11028 binary has a relatively large orbital
separation, our analysis of another wide binary in M67
\citep[WOCS 12009; ][]{s1247} indicated that the brighter star was likely the product
of an earlier merger. A key piece of evidence for that binary was the lack of
detectable Li. Single main sequence stars of similar brightness in the
cluster inhabit the Li plateau, a broad maximum in the Li
abundance. The lack of detectable Li was an indication of nuclear
processing, consistent with surface material having originally been in a lower-mass
main sequence star with a deeper surface convection zone.

To have greater assurance that the stars of WOCS 11028 have evolved as
isolated single stars and can be used to constrain the cluster age, we
can also use the Li abundance here.  If the brighter star in WOCS
11028 was unmodified since the cluster's formation, its mass
($\sim1.2\msun$) would place it toward the bright end of the cluster's
Li plateau with abundances $A(\mbox{Li}) \approx 2.5$ \citep{pace}. On
the other hand, the secondary star (mass near $0.9\msun$) should
not have detectable Li. Stars in the plateau have outer convective
zones that do not transport Li nuclei down to temperatures where nuclear
reactions can burn them, and additional mixing processes must have
little, if any effect. Even with the diluting effects of the secondary
star's light on the Li lines, Li should be detectable if it is present
with the plateau abundance.

\begin{figure}
\includegraphics[scale=0.55]{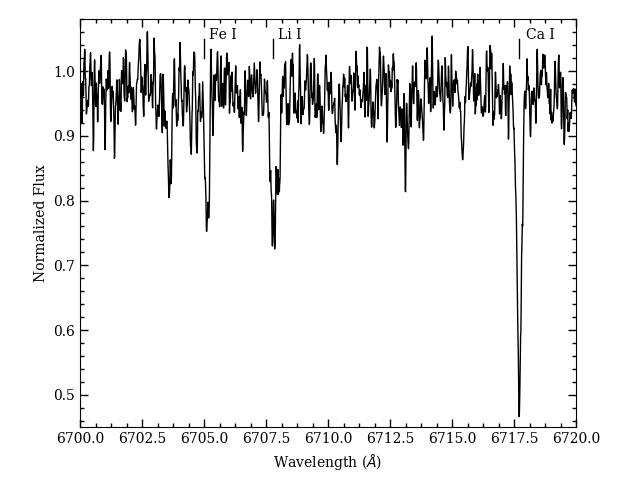}
\caption{A HARPS-N spectrum for WOCS 11028 in the vicinity of the Li I
  resonance doublet, shifted to account for the primary star
  velocity. \label{lispec}}
\end{figure}

We are able to detect Li in our HARPS-N spectra
as shown in Fig. \ref{lispec}. If the star had a history of
interaction (mass transfer, or a merger involving lower mass stars),
it is unlikely that the Li would be detectable. If the brighter star
formed in a merger, the more massive merging star could not have been
more than about $0.2 \msun$ less massive than the current star or else
a surface convection zone would have had a chance to consume its Li
during its main sequence evolution. In addition, the less massive
merging star would take up residence in the core of the merger remnant
because of its lower entropy, rebooting its nuclear evolution with gas
having higher hydrogen abundance.  After some adjustment on a thermal
timescale, it would have been bluer than other cluster stars of the
same mass.  A precise determination of the Li abundance is complicated
by the secondary star's light, and is beyond the scope of this work.
However, the strong Li detection coupled with strong indications that
WOCS 11028 A has a color consistent with its fellow cluster members
effectively rules out anything except a very early merger and/or a
merger with a very low mass object. Effectively, this would make
today's star the same as an unmodified star of the same mass.

We do not see evidence of rotational spot modulation in the K2 light
curves due to spots, but our HARPS-N spectra allow us to constrain the
rotational speeds of the stars in the binary via the widths of the
broadening function peaks. We are unable to detect rotation with an
upper limit $v_{\rm rot} \sin i < 5$ km s$^{-1}$. If the primary
  was pseudo-synchronized, its rotation velocity would be around this
  limit (rotation period near 14 d). However, the timescale calculated
  for this \citep{hut} is far longer than the cluster age, thanks to
  the substantial separation even at periastron. Considered as a
  single star, the primary has probably spun down somewhat, although
  not to the same degree as lower mass stars with deeper surface
  convection zones. \citet{angus} calibrated their gyrochronological
  models against the rotation of asteroseismic targets in the {\it
    Kepler} field, and their sample probably gives the best indication
  of typical rotation rates for old stars more massive than the
  Sun. Their isochrones predict a rotation period near 16 day, and
  this would imply a rotation velocity below our ability to detect it
  spectroscopically.
As an old cluster, M67 is expected to have slowly
rotating stars, but this does put a limit on how long ago mass
transfer or a merger could have happened if either star was somehow
modified. A rapidly rotating star with the mass of our primary star
would require approximately 3 Gyr to reach our detection limit
according to the rotational isochrones of \citet{angus}, and this is
the majority of the cluster's age.

\subsection{Binary Star Modeling}\label{binary}

We used the ELC code \citep{elc} to simultaneously model the
ground-based radial velocities and \kep K2 photometry for the WOCS
11028 binary.  The code is able to use a variety of optimizers to
search the complex multi-dimensional space of parameters used in the
binary star model.  In our particular case, we used a set of 15
parameters. Two of the parameters were the orbital period $P$, and the
reference time of periastron $t_P$.
Six additional parameters mostly characterize the orbit: the
velocity semi-amplitude of the primary star $K_1$, mass ratio $q=M_2 /
M_1=K_1/K_2$, systematic radial velocities\footnote{We allow for the
  possibility of differences for the two stars that could result from
  differences in convective blueshifts or gravitational redshifts.}
$\gamma_1$ and $\gamma_2$, eccentricity ($e$), and argument of
periastron ($\omega$).  For a typical double-lined spectroscopic and
eclipsing binary, both the radial velocities and eclipse light curves
constrain $e$ and $\omega$, with the phase spacing of the eclipse
playing a large role.  But because this binary only has one eclipse
per cycle, these parameters are constrained by the radial velocities
almost exclusively.

The light curve has a large role in constraining the inclination
parameter $i$, but there are potentially correlations between inclination
and choices for radius and temperature parameters. For radii, we used
the sum of the stellar radii $R_1 + R_2$ and their ratio $R_1 / R_2$
as parameters. For this binary, the sum of radii is constrained
by the measurements of the single grazing eclipse in each cycle along
with the well-determined spectroscopic parameters of the orbit. The
radius ratio can be constrained by luminosity ratios we have derived from
our SED analysis (\S \ref{sedsec}) in concert with temperature
information.  For a typical eclipsing binary, the temperature ratio
would be constrained by the relative depths of the eclipses, but that
is not measurable here. We have separate constraints
on the temperatures of the stars from SED fits to the proxy stars, and
so we use these as fit parameters: temperatures of the primary $T_1$
and secondary $T_2$.

Finally, there is some dependence of the fits on the limb darkening,
although relatively little because only the near-limb of the secondary
star is probed by the grazing eclipses we see.  So we fit for two
quadratic limb darkening law coefficients ($q_{1B}, q_{2B}$) for the
secondary star (the only star that is eclipsed) in the \kep bandpass.
Our runs of the ELC code employed the \citet{kipping} algorithm, which
involves a search over a triangular area of parameter space of
physically realistic values.  This forces the model star to a) darken
toward the limb, and b) have a concave-down darkening curve.
In the final analysis, we find that the limb darkening coefficients
are virtually unconstrained, but by allowing the code to
systematically explore potential values, the uncertainties will be
incorporated in our uncertainties of the other stellar parameters.  Owing to
the substantial orbital separation, we assumed that the stars are
spherical.  Given this, we used the algorithm given in \citet{short}
as implemented in ELC to rapidly compute the eclipses.  Finally,
because there is little or no out-of-eclipse light modulation, we did
not see a need to model spots.

For the K2 light curve modeling, we only included observations that
were within about $\pm0.01$ in phase of the eclipse, and a similar
section around where the other eclipse would have been if the binary
was closer to edge-on. In this way, our models ``see'' that there is
only one eclipse per orbital cycle. In doing this, we leave out the
majority of observations, which may be affected by systematics due to
spacecraft pointing corrections and other instrumental
signals. Because the K2 light curves use long cadence data with 30
minute exposures, they also effectively measure the average flux
during that time.  To account for this, the ELC code
integrates the computed light curves in each observed exposure window.

The quality of the model fit was quantified by an overall $\chi^2$
derived from comparing the radial velocity and light curve data to the
models, as well as from how well {\it a priori} constraints were
matched. For this binary, we used stellar temperatures derived from
SED fits and luminosity ratios in $BVI_CJHK_s$ derived from the
modeling of the binary SED with single cluster stars. To illustrate
the effect of these constraints, we conducted a run without the
luminosity ratios, as shown in Table \ref{parmtab}.

To try to ensure that different datasets were given appropriate
weights in the models, we empirically scaled uncertainties on
different datasets to produce a reduced $\chi^2$ of 1 relative to a
best-fit model for that particular dataset. For spectroscopic velocity
measurements, this meant scaling differently according to the source
spectrograph and the star (primary and secondary). The K2 photometric
light curve uncertainties were scaled independently.

\begin{figure}
\epsscale{1.3}
% clusters/m67/s617/elc/everest/redo/posterior.py
\plotone{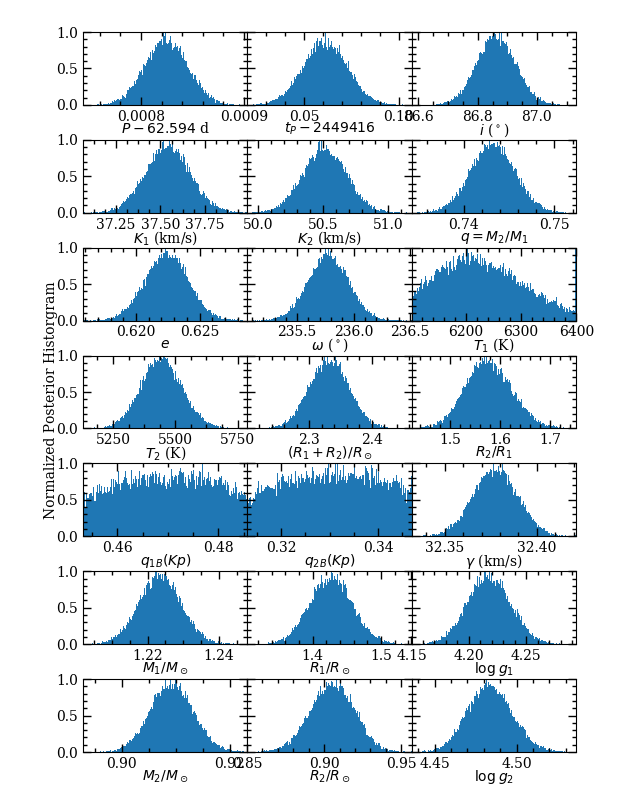}
\caption{Posterior distributions (normalized to the peak bin) for the
  fitted and calculated parameters for the EVEREST K2 data with SED
  luminosity constraints.
\label{posts}}
\end{figure}

\begin{figure}
\epsscale{1.3}
% clusters/m67/s617/elc/eclsphased.py
\plotone{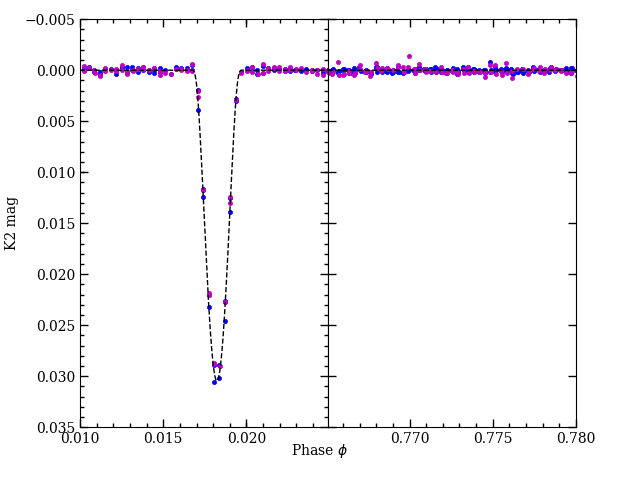}
\caption{K2 eclipse photometry (K2SFF pipeline in magenta, and EVEREST
  pipeline in blue) compared with the best-fit binary models (for the
  EVEREST pipeline). Orbital phases near the non-eclipsing conjunction
  are shown in the right panel. \label{phot}}
\end{figure}

We used a differential evolution Markov Chain optimizer \citep{demc} for
seeking the overall best-fit model and exploring parameter space
around that model to generate a posterior probability sampling.
Approximate $1\sigma$ parameter uncertainties were derived from the
parts of the posterior distributions containing 68.2\% of the
remaining models from the Markov chains.  Gaussians were good
approximations to the posterior distributions for all parameters
except limb darkening coefficients in the run where luminosity
constraints were applied.  The results of the parameter fits are
provided in Table \ref{parmtab}, and the posterior distributions
are shown in Figure \ref{posts}. A comparison of the K2 light curve
with the best-fit model is shown in Figure \ref{phot}, and a
comparison of the radial velocity measurements with models are shown
in Figure \ref{rvplot}.

\begin{deluxetable}{lccc}
\tablewidth{0pt}
\tabletypesize{\scriptsize}
\tablecaption{Best-Fit Model Parameters for WOCS 11028}
\label{parmtab}
\tablehead{\colhead{Parameter} & \colhead{No SED Constraint} & \multicolumn{2}{c}{SED Luminosity Constraints}\\
\colhead{Pipeline} & \colhead{K2SFF} & \colhead{K2SFF} & \colhead{EVEREST}}
\startdata
Constraints: & & & \\
$T_1$ (K) & $6200\pm100$ & $6200\pm100$ & $6200\pm100$ \\
$T_2$ (K) & $5450\pm100$ & $5450\pm100$ & $5450\pm100$ \\
%              MS fit                star fit    constraints
$L_2/L_1(B)$ & & $0.186\pm0.030$ & $0.186\pm0.030$ \\
$L_2/L_1(V)$ & & $0.227\pm0.030$ & $0.227\pm0.030$ \\
$L_2/L_1(I_C)$ & & $0.267\pm0.030$ & $0.267\pm0.030$\\
$L_2/L_1(J)$ & & $0.306\pm0.030$ & $0.306\pm0.030$\\
$L_2/L_1(H)$ & & $0.343\pm0.030$ & $0.343\pm0.030$\\
$L_2/L_1(K)$ & & $0.349\pm0.030$ & $0.349\pm0.030$\\
%$x_{K,1}$ & 0.454 & & $0.765\pm0.016$ \\%*
%$y_{K,1}$ & 0.271 & & $0.164\pm0.012$ \\%
%$x_{K,2}$ & $0.214\pm0.008$ & & $0.374\pm0.009$ \\
%$y_{K,2}$ & $0.748\pm0.035$ & & $0.771\pm0.014$ \\%*
\hline 
$\gamma$ (km s$^{-1}$) & $32.38\pm0.02$ & $32.37\pm0.02$ & $32.37\pm0.02$\\
$P$ (d) & 62.59484 & 62.59484 & 62.59482\\ 
$\sigma_P$ (d) & 0.00003 & 0.00003 & 0.00002\\
$t_P$ & 49416.060 & 49416.058 & 49416.061\\
$\sigma_{t_P}$ & 0.013 & 0.012 & 0.012\\
$i$ ($\degr$) & $86.70\pm0.15$ & $86.84\pm0.07$ & $86.86\pm0.07$ \\
$q$ & $0.743\pm0.003$ & $0.743\pm0.002$ & $0.743\pm0.003$\\
$e$ & $0.6227\pm0.0019$ & $0.6225\pm0.0017$ & $0.6222\pm0.0017$\\
$\omega$ ($\degr$) & $235.82\pm0.18$ & $235.70\pm0.18$ & $235.78\pm0.18$\\
$K_1$ (km s$^{-1}$) & $37.54\pm0.14$ & $37.54\pm0.13$ & $37.55\pm0.13$\\
$K_2$ (km s$^{-1}$) & $50.49\pm0.18$ & $50.51\pm0.18$ & $50.48\pm0.17$\\
$R_1/R_2$ & & $1.58\pm0.05$ & $1.57\pm0.05$\\
$(R_1+R_2)/\rsun$ & $2.37\pm0.07$ & $2.34\pm0.03$ & $2.34\pm0.03$\\
\hline
$M_1/\msun$ & $1.225\pm0.006$ & $1.222\pm0.006$ & $1.222\pm0.006$\\
$M_2/\msun$ & $0.911\pm0.004$ & $0.909\pm0.004$ & $0.909\pm0.004$\\
$R_1/\rsun$ & $1.44^{+0.08}_{-0.10}$ & $1.43\pm0.03$ & $1.43\pm0.03$\\
$R_2/\rsun$ & $0.86^{+0.15}_{-0.11}$ & $0.904\pm0.015$ & $0.908\pm0.015$\\
$\log g_1$ (cgs) & $4.21^{+0.07}_{-0.05}$ & $4.209\pm0.019$ & $4.217\pm0.019$\\
$\log g_2$ (cgs) & $4.51^{+0.10}_{-0.12}$ & $4.485\pm0.014$ & $4.481\pm0.014$\\
\enddata
\label{parmtab}
\end{deluxetable}

As can be expected, when we do not enforce luminosity ratio
constraints from SED fitting, there is no information on the ratio of
radii in the data, and as a result, the radii of the stars are very
poorly constrained. The mass determinations are essentially
unaffected, however, because that information is contained in the
radial velocity curves and the orbital inclination (derived from the
light curve).

\section{Discussion}\label{discuss}

M67 has long been studied because of similarities in chemical
composition and age with the Sun, but stars at the cluster turnoff are
approximately 20-30\% more massive than the Sun, with substantial
differences in internal physics.  Stars with masses of $1.1-1.3 \msun$
may generate convective cores late in their main sequence evolution,
and this plays a critical role in their evolution.

In the discussion below, we will leverage new information from two
sources --- precise measurements of the stars in the WOCS 11028
binary, and additional characterization of stars around the turnoff of
the cluster --- to identify conflicts with models, and to try to
understand the physics that might be causing those conflicts.

\begin{figure*}
% clusters/m67/sss/plotisogaiadm_zs.py
\includegraphics[scale=1.0]{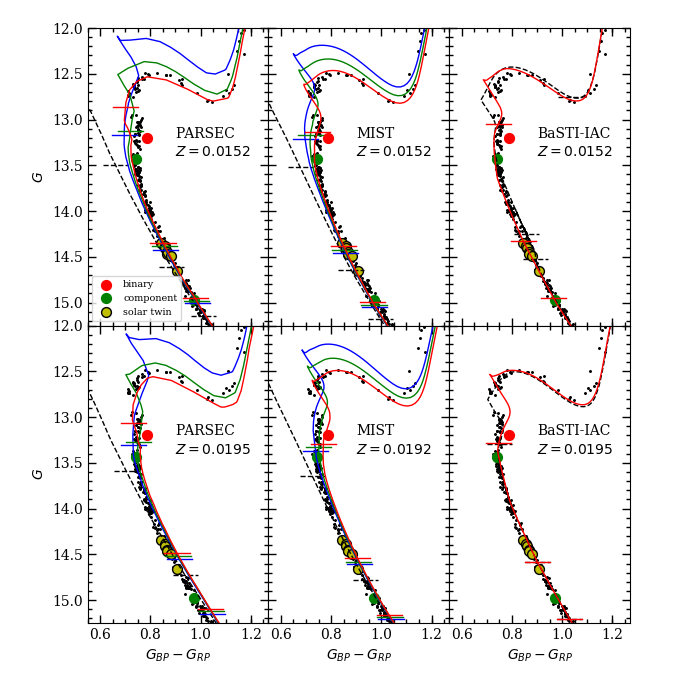}
\caption{{\it Gaia} color-magnitude diagram for the turnoff and main
  sequence of M67. Small points are probable single-star members and
  yellow circles are solar analogs identified by \citet{pasquini}. The
  system photometry for WOCS 11028 is shown with a red circle, and
  deconvolved photometry of the component stars are green
  circles. Theoretical isochrones have been shifted according to a
  corrected Gaia parallax [$(m-M)_0 = 9.63$], and reddening $E(B-V)=0.041$
\citep{taylor}. 
  PARSEC and MIST isochrones have ages 1.0 (dashed line), 3.2, 3.6, and 4.0 Gyr (solid blue, green, and red lines).
  BaSTI-IAC plots show an
 isochrone with no overshooting or diffusion (black dashed line), and
 with overshooting (solid red line) for 4.2 Gyr (top, right) and 4.0
 Gyr age (bottom, right).  The theoretical predictions for stars with
 masses equal to the primary star of WOCS 11028, the Sun, and the
 secondary star of WOCS 11028 are shown with horizontal
 lines. \label{gaiadmcmd}}
\end{figure*}

For clarity in the discussion, we define the upper turnoff to be at
the global minimum in color at the blue end of the subgiant branch ($G
\approx 12.74$); the turnoff gap slightly fainter than this ($12.75
\la G \la 12.95$); and the lower turnoff to be at the fainter local
color minimum ($G \approx 13.45$). Because WOCS 11028 A is nearly
exactly at the lower turnoff based on our SED decomposition (see
Figure \ref{gaiadmcmd}), it provides an excellent mass calibration
point for the cluster.

\subsection{The Cluster Composition}\label{m67comp}

As is usually the case, the chemical composition is an important
component of interpreting the measurements and understanding the
implications for the cluster age and for the stellar physics in the
models we use. We discussed high-precision cluster [Fe/H] measurements
relative to the Sun in \citet{s1247} in order to minimize systematic
errors due to composition. Since that paper though, it has become
clearer that diffusion is playing a role in the measured surface
compositions of M67 stars. Significant trends in abundance are seen as
a function of the evolutionary state of the stars, with subgiants
having higher measured abundances than turnoff stars
\citep{onehag,b_m,liu,souto19}. Because subgiants should have deep
enough convection zones to homogenize surface layers that were
affected by heavy element settling, the subgiant abundances should be
more representative of the initial bulk heavy element composition of
the stars. The spectroscopic studies generally come to the conclusion
that initial composition of the cluster members was
[Fe/H]$=+0.05$ to $+0.10$. In the model comparisons below, we will consider
initial metal compositions between solar and [Fe/H]$=+0.10$.

\subsection{The Characteristics and Evolution Status of WOCS 11028 A and B}\label{evol}

\begin{figure*}
% clusters/m67/s617/mesa/work/plottrackexs.py
\plotone{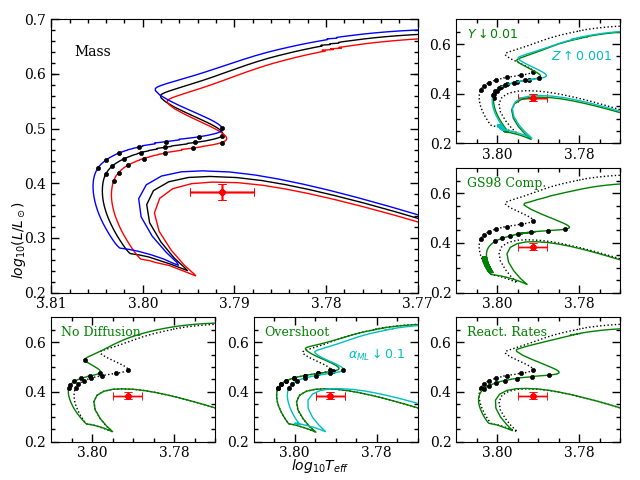}
\caption{Evolutionary tracks for WOCS 11028 A in the HR diagram using
  the MESA code. The red point shows the properties of WOCS 11028
  A. In most plots, the dotted line track is the baseline model shown
  in the upper left panel ($M=1.222 \msun$, $Z=0.0162$, \citealt{asplund} composition mix), and the colored tracks illustrate
    changes to the model physics.  Blue and red tracks in the upper
    right panel illustrate $1\sigma$ differences in mass from the best
    fit value. Black points show ages from 2.5 to 4 Gyr in intervals
  of 0.25 Gyr.
\label{hrtracks}}
\end{figure*}

The verticality of the cluster CMD at the position of WOCS 11028 A
constrains the present {\it direction} of its evolution track, and we
illustrate the discussion with MESA model tracks (version r11701;
\citealt{paxton11,paxton13,paxton15,paxton18,paxton19}) in the largest
panel of Figure \ref{hrtracks}. The models predict that for the first
2 Gyr of the star's life, it was evolving mostly upward in luminosity, with
slowly increasing temperature before hitting a maximum and starting to
decrease. The star must have started evolving more rapidly toward
cooler temperature in order to change the local slope of the main
sequence isochrone from increasing temperature to constant temperature with
increasing mass.  Comparing the markers showing common age for three
stars of slightly different mass, the triplet becomes vertical near
the midpoints of the tracks during this coolward movement. For greater
age, the triplets have decreasing temperature with higher mass, and in
M67 this corresponds to stars between the lower turnoff and the faint
end of the gap.

The precise point where the vertical slope is produced will depend on
details of the fuel consumption in this phase, but we can be assured
that the star must still be evolving faster toward lower temperature
than slightly lower mass stars (in other words, {\it accelerating}) or
else it would not be able to match the temperature of those stars that
should be evolving at a very similar speed. This in turn assures us
that convection must be present in the core of the star --- low-mass
stars that do not establish convective cores on the main sequence only
accelerate toward lower temperature when they are past central
hydrogen exhaustion while on the subgant branch. So the mass
measurement for WOCS 11028 A gives us an observational lower limit for
stars that establish convective cores on the main sequence.

In the evolution tracks, the turn toward lower temperature occurs when
the extent of the convective core increases in the latter half of core
hydrogen burning. The CNO cycle overtakes the $pp$ chain as the
dominant energy generating reaction network while central temperature
increases and hydrogen abundance declines.  So not only do we have
good measurements of the WOCS 11028 A star (described below), but we
have an indication of physical conditions occuring within the star. A
complete model of the star would match all of the measurements and
evolutionary constraints with a unique set of parameters, while
mismatches would indicate systematic errors.

\subsubsection{Mass and Radius}\label{mrsec}

Although the radii we derive from our analysis of the binary are not
as precise as they could have been if the system had been doubly
eclipsing, they can still provide some indication of the age
independent of other stars in the cluster if the stars have not
undergone mass transfer or mergers in their history\footnote{The more
  massive star in the binary WOCS 12009, analyzed in \citet{s1247} and
  mentioned in section \ref{liline}, had a much smaller radius than
  expected for reasonable cluster ages, providing support for the idea
  of a stellar merger in that case. The radius of the less massive
  star may not have been involved in the merger, but its radius is
  also unlikely to be significantly different even if it was.}. In
addition, these comparisons are independent of uncertainties in
quantities like distance and reddening.

\begin{figure}
\epsscale{1.3}
% clusters/m67/s617/plotmriso.py
\plotone{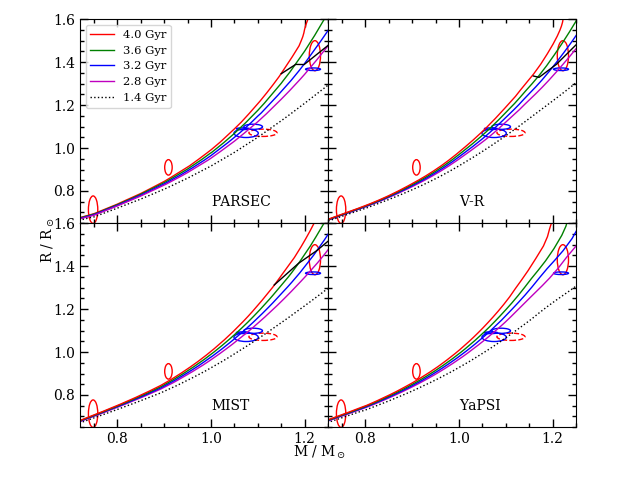}
\caption{Mass-radius plot for measured members of M67 (WOCS 11028 A and B, and WOCS 12009 B; red), with
  $2\sigma$ uncertainties indicated by the error ellipses. The
  probable straggler WOCS 12009 A is marked with a red dashed ellipse,
  and measured eclipsing stars in the younger cluster NGC 6819 are
  blue ellipses. Models use $Z = 0.0152$, 0.0188, 0.0142, and 0.0162,
  respectively for PARSEC \citep{parsec}, Victoria-Regina
  \citep{vandenberg2006}, MIST \citep{mist0,mist1}, and YaPSI
  \citep{yapsi} isochrones.
The solid black lines connect the lower turnoff points on the
different isochrones.
\label{tomr}}
\end{figure}

The comparison with several sets of solar-metallicity isochrone models
is shown in Figure \ref{tomr}.  Taken at face value, WOCS 11028 A is
smaller than predicted for typically quoted M67 ages ($3.5-4$ Gyr) by
$2\sigma$ or more, and WOCS 11028 B is larger than reasonable models
by a similar amount. Note that if we are systematically
underestimating the luminosity ratio used in our binary star modeling
(as one might wonder from the discussion of the SED fits in \S
\ref{sedsec}), this would exacerbate the conflict with models by
decreasing the primary star radius and increasing the secondary star
radius.

\begin{figure}
\epsscale{1.3}
% papers/m67/plotmrdebcat.py
\plotone{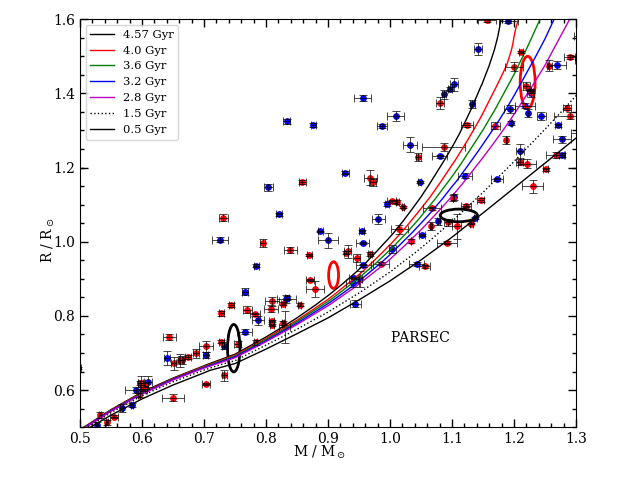}
\caption{Mass-radius plot for measured members of M67 (WOCS 11028,
  red; WOCS 12009 black) with $2\sigma$ uncertainties indicated by the
  error ellipses. Precisely measured eclipsing binary stars
  \citep[DEBCat, retrieved 2019;][]{debcat} are shown with $2\sigma$
  error bars (primary stars blue, secondary stars red). PARSEC
  \citep{parsec} isochrones are shown for $Z = 0.0152$. 
\label{mrdeb}}
\end{figure}

When compared with stars from well-measured binaries
(\citealt{debcat}; see Figure \ref{mrdeb}), the
secondary stars of the two measured M67 binaries (WOCS 11028 B and
WOCS 12009 B) do not look out of the ordinary. 
PARSEC models {\it generally} underpredict radii for stars
with $0.6 \le M/\msun \le 0.95$ like WOCS 11028 B, and this is not
sensitive to the assumed age or metal content $Z$.
The disagreement with models leads to some
justifiable concerns whether a model age can be trusted if the
characteristics of the relatively unevolved stars cannot be predicted
accurately.

Among the stars in the \citet{debcat} sample are several from the
younger but similarly metal-rich cluster NGC 6819 \citep[$\sim2.2$
  Gyr;][]{brewer}.  One of the stars is WOCS 40007 A, which has a mass
($1.218\pm0.008\msun$) that is nearly identical to that of WOCS 11028
A. As another indicator of M67's relative age, WOCS 11028 A is at
M67's turnoff, while WOCS 40007 A is approximately 0.85 mag fainter
than NGC 6819's turnoff in $V$. WOCS 40007 A has a radius
($1.367\pm0.003\rsun$) that is smaller than WOCS 11028 A, although
WOCS 11028 A's radius is measured with about 10 times lower
precision. So while WOCS 11028 A appears to be somewhat small relative
to the model expectations for M67's age, it does appear to be larger
and older (by about 0.4 Gyr) than the corresponding stars of NGC 6819.

An absolute determination of the age requires comparisons to models
though, and the characteristics of WOCS 11028 A return a value of
$3.0\pm0.3$ Gyr for isochrones having $Z\approx0.015$. Radius is
fairly insensitive to one of the more important model uncertainties
--- bulk metal content $Z$ --- and so should be given extra attention
in considering the age. From experiments with MESA models, we find
that $R \propto Z^{-1/4}$ for stars near WOCS 11028 A's mass, whereas
$L \propto Z^{-1}$. By the same token, helium abundance $Y$ generally
affects radius in the opposite direction, and if one assumes, as most
isochrone sets do, that galactic chemical evolution increases $Y$ and
$Z$ together (in other words, $\Delta Y / \Delta Z$ is a constant),
radius appears even less sensitive to composition changes. We will return
to this point in Section \ref{tl}.

We should also consider the evolutionary
status of the brighter star WOCS 11028 A though. If models agree with our
mass-radius measurements at a particular age but have the star at the
wrong evolutionary state, it would be an indication of a systematic
error in model physics. In Figure \ref{tomr}, we have connected
isochrone points corresponding to the lower turnoff at each age. We
find that, for all of the model sets we could check, the mass-radius
combination for WOCS 11028 A is consistent with model predictions for
that evolutionary stage.

\subsubsection{The Color-Magnitude Diagram (CMD)}\label{cmd}

To date, the primary method for measuring the age of M67 has been
isochrone fits to the CMD.  Examples of results include 3.5 Gyr to 4.0
Gyr \citep{saraj} and 3.6 to 4.6 Gyr \citep{vsm67}. A large portion of
these age ranges result from model physics and chemical composition
uncertainties that impact this cluster, and new information may
help reduce the age uncertainty by reducing these modeling
uncertainties. For example, \citet{stellom67k2} used an asteroseismic
analysis of the masses of pulsating giants in M67 to derive an age of
about 3.5 Gyr, although that age still has uncertainties
due to model physics.

The modeling of the eclipsing binary star gives us very precise masses
for the component stars, and our SED analysis places
restrictions on the CMD positions of the stars. These are the most
precise data that can be extracted from the binary at present. When
utilized with precise photometry for other cluster members, isochrone
fitting can provide us with an estimate of the age. It is important to
remember that masses are generally unavailable for traditional CMD
isochrone fitting, and by employing them here, we can start to address
systematic uncertainties on the derived age.

The {\it Gaia} photometric dataset is one of the most precise
available, so in Figure \ref{gaiadmcmd} we compare the M67 photometry
with theoretical isochrones. The isochrones were shifted based on the
mean {\it Gaia} parallax for the cluster corrected for the
\citet{schonrich} zeropoint offset [for a distance modulus $(m-M)_0 =
  9.63$), along with the \citet{taylor} reddening for the cluster
  [giving extinction $A_G = 0.111$ and $E(G_{BP} - G_{RP}) = 0.060$].
  The comparison indicates that the position of the faint component of
  WOCS 11028 is consistent with the
  theoretical predictions for its mass, especially considering an
  $1\sigma$ uncertainty in $G_B$ of about $\pm0.075$ mag (see
  section \ref{sedsec}). There is significant disagreement between the
  position of WOCS 11028 A and the model predictions for its mass,
  however.  The theoretical predictions are at least about 0.17 mag
  too bright for the youngest models under consideration, and closer
  to the brightness level of the {\it binary}. It should be remembered
  as a practical matter that the magnitude of the brighter star is
  much more certain than the fainter
  star because of the need to match the brightness constraint of the
  binary, assuming that both stars are radiating like normal single
  star cluster members.  Our earlier analysis puts the
  $1\sigma$ fit uncertainty at around $\pm0.018$ mag in $G$, so that
  an explanation beyond statistical uncertainty seems necessary.

The fact that model predictions fall near or brighter than the
combined brightness of the binary reduces the range of possible
explanations: either our assumption that the binary's individual stars are
emitting like unmodified single main sequence stars is wrong (and the
binary's light should not be modeled using stars on the main
sequence), or there is a significant systematic error with model
predictions. We have not seen any evidence to support the idea that
the primary star is different from similar main sequence stars: there
is detectable Li and the rotational velocity is low and consistent
with a normally-evolving single star (section \ref{liline}), and the
ultraviolet emission from the binary can be reproduced very well by
the emission of other cluster main sequence stars (section
\ref{sedspec} and Figure \ref{galspec}). If the primary star had been
modified earlier in its history, it should appear fainter and bluer
than single main sequence stars that evolved uneventfully
with the cluster.

\begin{figure}
\epsscale{1.3}
% papers/m67/plotdebmags.py
\plotone{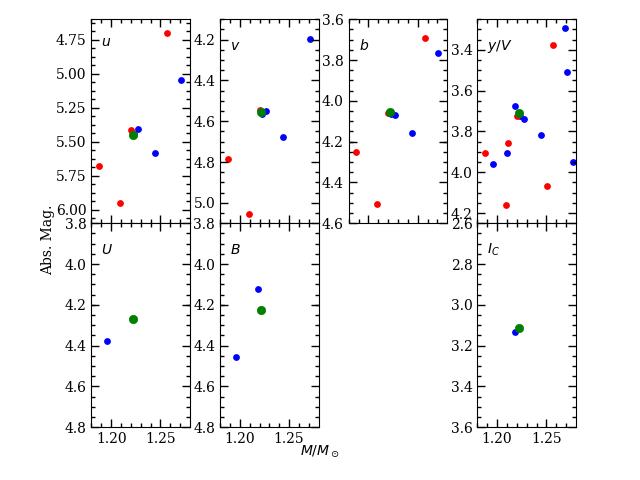}
\caption{Absolute magnitudes for main sequence stars with masses near
  that of WOCS 11028 A (green points). Blue points are primary stars of
  detached eclipsing binaries, and red points are secondary
  stars.\label{debmagfig}}
\end{figure}

We can also compare with stars from field eclipsing binaries having
well-measured luminosity ratios in common filter bands. For this
purpose, we examined stars from the DEBCat database \citep{debcat}
with stellar masses within about $0.05\msun$ of that of WOCS 11028 A,
and kept those stars with precise {\it Gaia} parallaxes.  The stars in our
comparison are given in Table \ref{debmags}, and a comparison of the
absolute magnitudes is shown in Figure \ref{debmagfig}. WOCS 11028 A
is generally at the bright end of the range covered by stars with the
most similar masses, which is to be expected given the age of the
cluster it resides in. Even so, its magnitudes are quite comparable to
several other stars: UX Men A and WZ Oph A and B agree extremely well
with WOCS 11028 A in Str\"{o}mgren filters, and WOCS 40007 A from the
cluster NGC 6819 agrees quite well in $BVI_C$ magnitudes if we make
use of a distance modulus derived from the eclipsing binaries in the
cluster (as it is too distant for a good {\it Gaia} parallax).  We
conclude that WOCS 11028 A is not unusually faint for its mass, and
this cannot explain the disagreement with model isochrones.

\subsubsection{Temperature and Luminosity}\label{tl}

\begin{figure}
\epsscale{1.3}
% clusters/m67/s617/plotmtiso.py
\plotone{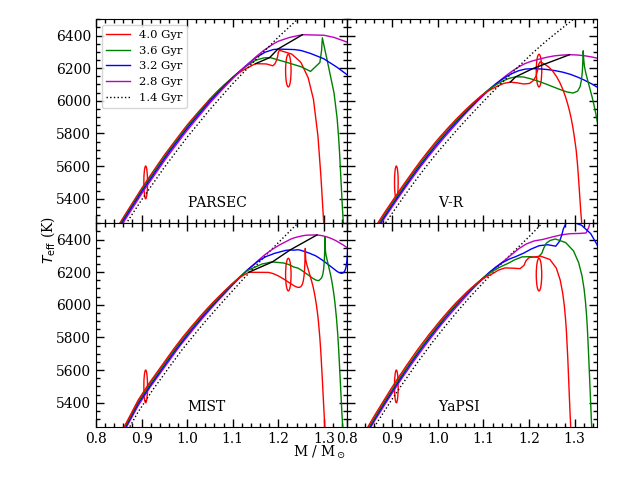}
\caption{Mass-temperature plot for measured members of M67 (red), with
  $2\sigma$ uncertainties indicated by the error ellipses. 
  Models use $Z = 0.0152$, 0.0188, 0.0142, and 0.0162, respectively for PARSEC
\citep{parsec}, Victoria-Regina \citep{vandenberg2006},
MIST \citep{mist0,mist1}, and YaPSI \citep{yapsi} isochrones.
The solid black lines connect the lower turnoff points on the different isochrones.
\label{tomt}}
\end{figure}

With the variety of photometric measurements made for M67 stars, the
SEDs provide us with excellent leverage on the temperatures of
the binary star components as well as other cluster stars for
comparisons. We find that both components of the binary agree well
with stars from the DEBCat database, and the temperature of the
primary star matches well with spectroscopic temperatures (see \S
\ref{sedspec}). While WOCS 11028 B matches the model temperature for
its mass (see Figure \ref{tomt}), WOCS 11028 A is generally around 100
K cooler than the model that has its lower turnoff at the appropriate
mass.

We can compute the luminosities for the stars in two different
ways. The first uses the bolometric flux derived from SED fitting
along with the {\it Gaia}-based distance and the inverse square
law. The bolometric flux is computed assuming that any flux that is
not covered by photometric measurements can be accounted for using an
appropriate synthetic spectrum. The luminosities are found to
be $2.43\pm0.08$ and $0.61\pm0.02 \lsun$ for the proxies of the two
stars in the binary. The quoted uncertainties include contributions
from the parallax and the bolometric flux fit.
Alternately, we can compute luminosities from the radii derived in
the binary star analysis and the temperatures derived from the SED
fits. Here the results are $2.72\pm0.22$ and $0.68\pm0.056 \lsun$,
respectively. These estimates are marginally consistent with the
values computed from the bolometric fluxes, but are less precise.

\begin{figure}
\epsscale{1.3}
%% clusters/m67/s617/mesa/work/plotmesachars.py
\plotone{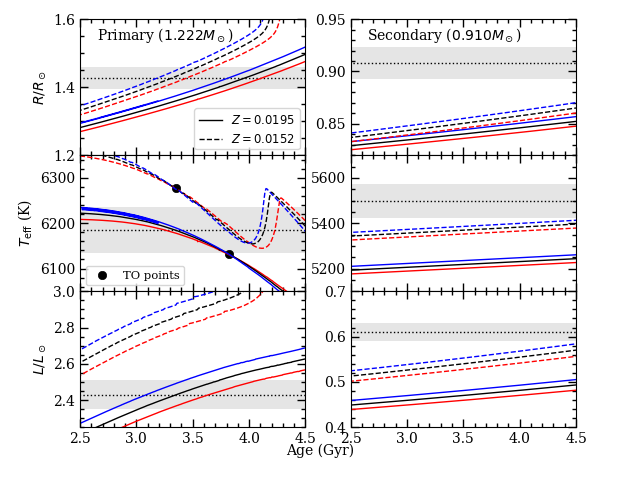}
\caption{Stellar characteristics versus age for MESA models. Gray
  boxes show $1\sigma$ uncertainties on the measured radius,
  temperature, and luminosity of WOCS 11028 A (left) and B
  (right). MESA evolution tracks for $Z = 0.0195$ (solid lines) and $Z
  = 0.0152$ (dashed lines) are shown, with masses $M = 1.222 \msun$
  (black) in the left column and $0.910 \msun$ in the right
  column. Tracks for masses $1 \sigma$ higher (blue) and lower (red)
  are also shown. Black points in the left middle panel show
    lower turnoff points (where stars of similar mass hit the same
    temperature).
\label{zage}}
\end{figure}

As seen in Figure \ref{zage}, WOCS 11028 B is in general agreement with model luminosity
predictions at solar metallicity, but WOCS 11028 A is about 25\%
lower than model values. This is a stronger statement than it might
seem at first because we have the additional constraint on the
evolutionary state of WOCS 11028 A. While it is possible to find a
match with the models in luminosity and mass, this occurs at young
ages when a star of WOCS 11028 A's mass is not precisely at the lower
temperature maximum at the cluster turnoff in the cluster CMD (see \S
\ref{evol}). The discrepancy is potentially a serious issue for age
determination because it implies a significant systematic
error. Below we discuss factors that impact the luminosity
disagreement, and the difficulty of finding a clear explanation.

For comparison, we show exploratory MESA evolution models in Figure
\ref{hrtracks}. For these experiments, the baseline inputs were a
solar-calibrated \citealt{asplund} composition ($Y = 0.2703$, $Z =
0.0162$) and mixing length parameter ($\alpha = 1.80$), nuclear
reaction rates from the JINA compilation, diffusion, and disabled core
convective overshooting. Before proceeding, we note that convection or
diffusion parameters produce minimal effects on the luminosity of the
star as it moves away from the main sequence toward its temporary
temperature minimum shortly before core hydrogen exhaustion. These factors play
more of a role when discussing the rest of the turnoff in \S
\ref{turnoff}.
\begin{itemize}
\item Measured mass: A stellar mass approximately $0.05\msun$ lower
  than measured would allow the redward evolution of the model star to
  pass through the middle of the HR diagram error bars. This would be
  a $10\sigma$ error according to our calculations, however, and seems
  unlikely. We note that a star of $1.17\msun$ still produces a
  significant convective core at the end of its main sequence
  evolution in the MESA models.
\item Measured luminosity: If there is a systematic error in our
  decomposition of the binary's light or in the distance of the
  system, it could account for some of the luminosity discrepancy. If
  the distance is underestimated (due to an overestimate in the
  parallax), the luminosity would be larger. The {\it Gaia} parallax
  correction would be a suspect because it is acting in this
  direction. The corrections being discussed (\S \ref{dm}) would
  produce a maximum effect of 0.04 dex in $\log_{10} (L / \lsun)$,
  which is not enough to explain the full discrepancy, and even so,
  smaller distances become more inconsistent with CMD-based distance
  measures. As for the bolometric flux, we note that the star-based
  decomposition of the SED (instead of the main sequence fit) leads
  to a lower $F_{\rm bol}$, and so would exacerbate the discrepancy if
  true.
\item Nuclear reaction rates: CNO cycle rates in a stellar evolution
  code affect the strength of the energy generation. Larger rates lead
  to an earlier initiation of the convective core and lower luminosity
  for the star during its redward evolution. The CNO cycle bottleneck
  reaction $^{14}$N(p,$\gamma$)$^{15}$O is the most important to
  consider. The reaction rate was revised downward by nearly 50\% at
  stellar energies as a consequence of a reduced contribution of capture
  to the ground state of $^{15}$O, making capture to the 6.792 MeV
  excited state the dominant channel.
Summaries of relevant experimental measurements can be found in
\citet{sf2} and \citet{wagner}.  The NACRE reaction rate tabulation
\citep{nacre} uses the older, higher astrophysical $S$ factor, while
recent tabulations \citep{nacre2,jina} use values consistent with the
recent experimental results \citep{formicola,im05}.  While there
appears to be convergence on experimental details of the
$^{14}$N(p,$\gamma$)$^{15}$O reaction, the extrapolation of the $S$
factor to stellar energies now contributes the most to the uncertainty
at the level of about 8\% \citep{wagner}. Although significant for
other purposes (such as using solar neutrinos to address solar
composition uncertainties), this level of uncertainty is too small to
account for the luminosity discrepancy unless there is an unknown and
more substantial systematic error in the model reaction rate.

The bottleneck reaction in CNO cycle II is $^{17}$O$(p,\alpha)^{14}$N,
and this rate has recently been revised upward by a factor of 2 at
stellar energies by \citet{brunoo17}. This affects the flow of
$^{16}$O into the main CNO cycle, and so has a small effect on the
luminosity. But again, within the likely uncertainties, reaction rates
also fall far short of explaining the size of the luminosity
discrepancy.
\item CNO and heavy element abundances: CNO element abundances affect
  the CNO cycle energy generation through their role as catalysts, and
  larger abundances allow the convective core to be established
  earlier in the main sequence evolution and at lower luminosity. As
  an illustration in the middle right panel of Figure \ref{hrtracks},
  the use of a GS98 solar abundance mix moves the redward evolution to
  lower luminosity because that mix has a larger proportion of CNO
  elements and because solar calibration with this composition leads
  to a larger metallicity. However, even a nearly
  25\% change in the CNO elements is not enough to explain the
  luminosity discrepancy for WOCS 11028 A.
As seen in the upper right panel of Figure \ref{hrtracks}, increased
metal content also results in a reduced luminosity during the redward
evolution. As discussed in section \ref{m67comp}, evidence from recent
spectroscopic studies suggests that subgiants (which should have deep
enough outer convective zones to erase diffusion effects that
occurred on the main sequence) have higher surface metal abundances than
turnoff stars.

Reasonable changes to the bulk heavy element abundance can bring the
models into agreement with the measured luminosity and temperature of
WOCS 11028 A at its measured mass. However, if we assume that
composition changes are responsible, then this affects the agreement
of other well-measured stars with the models. In particular, the
models become too low in luminosity to match WOCS 11028 B.
\end{itemize}

Our conclusion is that a larger bulk heavy element abundance for M67
stars (previously masked by diffusion effects) is the most significant
effect on the agreement between models and the characteristics of WOCS
11028 A that we can identify. Figure \ref{zage} compares the
characteristics of WOCS 11028 A with MESA models for $Z = 0.0152$ and
$0.0195$. In the $\teff$ plot, the faint cluster turnoff occurs at the
age where the three tracks cross, indicating a constant temperature in
the mass range shown. With the larger bulk metal content, the radius,
effective temperature, and luminosity of the star are approximately in
agreement with models for ages roughly between 3.5 and 4.0
Gyr. Further adjustment of $Z$ would throw either the model $\teff$ or
$L$ out of agreement with the observations, although the radius is
less sensitive to metal content.

If the same large metal abundance is applied to models of WOCS 11028
B, they are pushed farther from agreement with the observations,
however. The radius in particular should not be sensitive to the metal
content, but the models are significantly lower than the observations
for reasonable ages. As seen in Figure \ref{mrdeb}, the measurements
of WOCS 11028 B agree with those for other well-measured eclipsing
binary star components. As a result, we think that there are still
unidentified systematic errors that are playing a role. 

\subsection{The Turnoff of M67 and Convective Cores}\label{turnoff}

With the new data on the massive star in the WOCS 11028 binary, there
is an opportunity to re-examine the cluster turnoff and the
constraints it places on the age and stellar physics. We can make a
more precise estimate of star masses at the turnoff than has been
possible before by employing models that are capable of reproducing
the characteristics of WOCS 11028 A, reducing the importance of
systematic errors by making a {\it relative} comparison. The observed
properties of the CMD gap and the upper turnoff are strongly
influenced by the extent of the convective core, which in turn affects the 
amount of accessible fuel and the main sequence lifetimes.

A major uncertainty about convective core extent is overshooting
  of the convective boundary. The stars at M67's turnoff are uniquely
  interesting because they are close to the minimum mass for having
  convective cores, and the amount of overshooting could play an
  outsized role in stars with small cores.  Some studies of eclipsing
  binaries in the $1.2-2.0 \msun$ mass range \citep{ct17,ct18}
  indicate that convective overshooting is consistent with zero at the
  mass of WOCS 11028 A, but rapidly increasing to a plateau for masses
  greater than about $2 \msun$.  On the other hand, the models of
  \citet{higl} also show little or no need for overshooting in
  binaries with main sequence stars having masses near $1.2 \msun$,
  but some weak evidence for overshooting in BG Ind, having a somewhat
  more massive ($1.42 \msun$) and more evolved primary
  star. \citet{const} found little evidence for a mass-dependent
  overshooting for masses between 1.2 and $2.0 \msun$.  A major
  problem with the binary star samples in most of these studies is
  that overshooting reveals itself most clearly very close to core
  hydrogen exhaustion, and unless stars are in a very specific
  evolutionary phase, their conventional stellar properties (like $R$,
  $L$, and $\teff$) will not be sensitive to the overshooting.  This
  disagreement in the literature reinforces the value of using the
  information contained in M67, with its larger sample of stars and mass 
inferences from the WOCS 11028 binary.

More recently, a number of asteroseismic studies have also
  attempted to constrain convective core overshooting with low-degree
  p modes. Studies like \citet{viani} find positive trends in
  overshooting amount with mass, while others like \citet{angelou}
  find overshooting amounts that are consistent with what is found
  from binary and cluster calibrations but without evidence of a clear
  mass dependence. These issues highlight that there are systematics
  and degeneracies in the fitting of parameters in asteroseismic
  studies, and that they have not fully clarified the overshoot issue.
  Asteroseismic masses have significantly lower precisions than binary
  system masses as well. All of these factors call out for additional
precise constraints on the overshooting in this mass range.

Overshooting did not play a significant role in our interpretation of
the WOCS 11028 binary because the primary star is well before core
hydrogen exhaustion, but it does more strongly affect the CMD
positions of stars around the turnoff gap and beginning of the
subgiant branch. 
Models indicate that a star like WOCS 11028 A reaches a maximum in
surface temperature when the convective core starts expanding as the
core temperature increases and the CNO cycle becomes more dominant.
After the star passes its surface temperature maximum, models show
that the convective core is maintained and the star evolves mostly in
$\teff$ until shortly before core hydrogen exhaustion. At that time
the convective core rapidly shrinks and disappears, and the star
executes a quick movement to higher temperatures and luminosities in
the HR Diagram. There is still a residue of hydrogen in the core and a
steep gradient in hydrogen abundance outside that, so the evolution of
the star slows down again in the early subgiant branch. Overall, this
evolution track is a hybrid, with the early core hydrogen burning
evolution similar to low-mass stars, and the late evolution similar to
more massive stars.

The extent and lifetime of the convective core in stars slightly more
massive than WOCS 11028 A is imprinted on the shape of the turnoff in
the CMD and in the number of stars present. The noticeable gap seen at
$12.7 \la G \la 13$ in M67 is a demonstration of the rapid evolution that results
from the disappearance of the convective core shortly before central
hydrogen exhaustion. Similarly, the fairly large density of stars
present just brighter than the gap shows that gas
with large hydrogen abundance is residing just outside the exhausted
center of the star, and the burning of this hydrogen allows the evolution to
slow down again.

To define the shape of the turnoff, we derived bolometric fluxes
and surface temperatures from SED fits for probable single stars at
critical points in the CMD. We identified WOCS 7006 / Sanders 1003 as
a likely single cluster member at the bright end of the gap based on
CMD position and lack of radial velocity variation \citep{geller}, and
WOCS 8011 / Sanders 1219 as a likely single member at the faint
end. We also examined WOCS 3003 / Sanders 1034 as a
representative of the end of the most heavily populated part of the
early subgiant branch.  The stars are shown in CMDs in Figure
\ref{turnoffs}, and their derived properties are given in Table
\ref{eepstars}.

\begin{deluxetable*}{lccccc}
\tablewidth{0pt}
\tabletypesize{\scriptsize}
\tablecaption{SED Fit Results for Turnoff Stars}
\label{eepstars}
\tablehead{\colhead{} & \colhead{WOCS 3003} & \colhead{WOCS 7006} & \colhead{WOCS 8011} & \colhead{WOCS 11028A} & \colhead{WOCS 11028B}}
%                              S1034            S1003                 S1219
\startdata
$F_{bol}$ ($10^{-11}$ erg cm$^{-2}$ s$^{-1}$) & $25.65^{+0.04}_{-0.23}$ & $21.51\pm0.07$ & $16.23\pm0.08$ & $10.89\pm0.03$ & $2.73\pm0.02$\\
$L$ ($\lsun$) & $5.70\pm0.20$ & $4.78\pm0.16$ & $3.62\pm0.12$ & $2.42\pm0.08$ & $0.68\pm0.06$\\
$V$ & $12.656\pm0.002$ & $12.824\pm0.002$ & $13.129\pm0.003$ & 13.568 & 15.167\\
$b-y$ & $0.381\pm0.001$ & $0.359\pm0.001$ & $0.376\pm0.001$ & 0.372 & 0.486\\
$c_1$ & $0.408\pm0.003$ & $0.440\pm0.004$ & $0.341\pm0.004$ & 0.394 & 0.321\\
H$\beta$ & $2.599\pm0.009$ & $2.611\pm0.008$ & $2.637\pm0.016$ & & \\
$T_{\rm eff}(b-y)$ (K) &  & 6225 & 6150 & 6225\\
$T_{\rm eff}$(IRFM) (K) & 6010 & 6170 & 6080 & 6185 & 5500\\
\enddata
\end{deluxetable*}

As was shown in Figure \ref{hrtracks}, several model physics parameters
play a role in setting the shape of the turnoff and subgiant branch, and most affect 
core convection. As a result, there is some degeneracy in the parameters that
can be used to model the turnoff. For our purposes here, we will focus on
convective core overshooting.
Overshooting has the most substantial effects on
evolution tracks shortly before the time of core hydrogen exhaustion.
Core hydrogen exhaustion is delayed by overshooting, which allows the redward
evolution to extend to lower temperature, giving isochrones a deeper
redward kink (see bottom middle panel of Figure \ref{hrtracks}). In
addition, the morphology of the upper turnoff changes in response to
overshooting, with increasing amounts resulting in a more horizontal
early subgiant branch. (See Figure \ref{gaiadmcmd}, and compare the
BaSTI-IAC isochrones with and without overshooting, or younger PARSEC
isochrones to older ones that have less overshoot.) Both the
morphology of the upper turnoff and the luminosity and temperature
positions of the stars measured near the gap in M67 are indicative of
weaker convective overshooting than is used in the MIST
isochrones. This is a means of calibrating overshoot for stars
with masses around the gap (approximately $1.32\msun$, based on models
that match WOCS 11028 A). The exact amount depends somewhat on other
physics, like diffusion rates, CNO cycle reaction rates, and CNO element abundances.

It has long been known that the turnoff of M67 is not well
matched by isochrones \citep{vg07,magic}.
Currently published theoretical isochrones should not be used to
calibrate the overshooting or read an age because they do not
reproduce the characteristics of WOCS 11028 A (see Section \ref{tl}),
and would therefore have clear systematic errors. There are
unfortunately several composition and physics factors that can affect
the size (and even the presence) of a convective core in models, and
relatively small changes in the extent of this core can strongly
affect the amount of hydrogen fuel the core gets, and therefore, the
length of the star's life. But by using models that are at least able
to match WOCS 11028 A, we should be able to reduce the size of the
errors.

\begin{figure}
% clusters/m67/s617/mesa/work/plotmesaseq.temp.py
\plotone{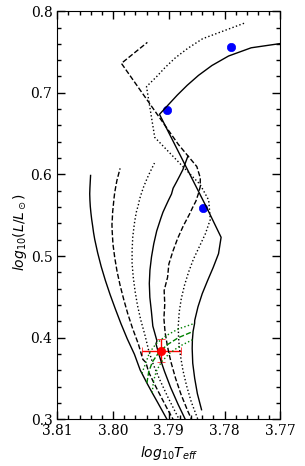}
\caption{Isochrones for ages between 2.5 and 4.0 Gyr (0.25 Gyr
  spacing) from MESA models for $Z = 0.0195$ and half strength
  convective core overshooting. The dashed green line connects models for the
  measured mass of WOCS 11028 A, with $\pm1\sigma$
  uncertainties. Measured characteristics of WOCS 11028 A (red) and
  stars at color-magnitude diagram turning points (blue) are shown
  with points. 
  \label{seqs}}
\end{figure}

Figure \ref{seqs} shows isochrones from MESA models having $Z=0.0195$ and
$Y = 0.2703$, with convective core overshooting set to half the
default value, or $f = 0.008$. Larger amounts of overshooting move the
gap to lower temperatures, and appear to be ruled out. With this set
of input parameters, we are able to approximately reproduce the HR
diagram positions of the three stars (described above) that reside at
identifiable points in the cluster CMD using an age around 3.75-4 Gyr. 
In addition, \citet{stellom67k2} found an average giant star mass
  of $1.36\pm0.01 \msun$ in their asteroseismic analysis of M67 giant
  stars using K2 data. This measurement is also consistent with
  an age near 4 Gyr for the MESA models used in Figure
  \ref{seqs}. The agreement with all of these observations implies
that the models are approximately reproducing the hydrogen abundance
profile in the stars at core exhaustion.

\section{Conclusions}

We present high-precision measurements of the masses, radii, and
photometry for stars in the bright detached eclipsing binary system
WOCS 11028 in the cluster M67. This cluster is commonly used as a
testbed for stellar evolution models of solar metallicity, and the
addition of new observables for stars in the cluster will further test
the fidelity of the physics we use to model them. 

Despite orbits that only produce one eclipse per cycle, we have shown
that high-precision masses for the stars in WOCS 11028 are
possible. We have used a decomposition of the SED of the binary to
constrain the temperatures, radii, and luminosities of the stars as
well. The brighter star in the binary sits precisely at the fainter
turnoff point for the cluster, where stars with convective cores are
accelerating toward the red in the later stages of core hydrogen
burning. The known mass of a star at an identifiable evolutionary
point constrains the age of the cluster, although uncertainties in
model physics reduce the precision that is attainable.

Systematic errors in the model physics remain a difficult problem for
stars at the turnoff of M67. Recent spectroscopic studies seem to make
a strong case for the inclusion of diffusion in models, and a larger
bulk metallicity for cluster stars can explain the characteristics of
WOCS 11028 A. However, the cost of this is a disgreement between
fainter stars (solar analogs and less massive stars) and models. We
are forced to the conclusion that there appears to be an unidentified
systematic error that affects the luminosities of stars between (and
possibly including) 1 and $1.2 \msun$. Our examination of physics
uncertainties like nuclear reaction rates, convection, and chemical
composition has not revealed an obvious culprit.
With the assumption of a larger bulk metallicity, age indicators are
somewhat consistent in returning ages between about 3.5 and 4
Gyr. Future adjustments to model physics will hopefully provide
greater agreement between the measures.

With a star of known mass at the lower cluster turnoff, we can more
reliably calibrate models of stars at core hydrogen exhaustion, which 
are sensitive to the size of the convective core. The morphology of
M67's turnoff implies that convective overshooting in turnoff mass
stars is present, but to an extent that is smaller
than used in standard isochrones.
After identifying composition and physics parameters that allow us to
match the characteristics of WOCS 11028 A, we can estimate the mass of
stars at the turnoff of M67. We find that stars near the gap and upper
turnoff should have masses of  $1.32\pm0.02 \msun$.
With these same model parameters, we can approximately match the
average asteroseismic giant mass of $1.36\pm0.01 \msun$ measured by
\citet{stellom67k2} at an age near 4 Gyr.  More precise comparisons
between observed masses and theory will require a better understanding of the
core convection physics because this critically affects the timing of
a star's divergence from the main sequence. Precise and direct
measurements of masses in the eclipsing system HV Cnc could improve
the comparison as well because the brightest star resides at the upper
turnoff. The difficulties in completing such a study are the faintness
of the companion and the contaminating effects of an unresolved third
star \citep{s986}. However, this is one of the few ways of further
identifying and eliminating the systematic errors in our modeling of
these stars and our determination of their ages.

\acknowledgments E.L.S. gratefully acknowledges support from the
National Science Foundation under grant AAG 1817217, and thanks
K. Brogaard for providing the original version of the spectral
separation code used in this work. R.D.M. acknowledges funding
support from NSF AST 1714506 and the Wisconsin Alumni Research
Foundation. D.S. is the recipient of an Australian Research Council
Future Fellowship (project number FT1400147).

This paper includes data collected by the {\it K2}
mission, and we gratefully acknowledge support from NASA under grant
NNX15AW69G to R.D.M. Funding for the {\it K2} mission is provided by
the NASA Science Mission Directorate.

This research made use of
the SIMBAD database, operated at CDS, Strasbourg, France;
the VizieR catalogue access tool, CDS, Strasbourg, France (DOI: 10.26093/cds/vizier; \citealt{vizier})
data provided by the High Energy Astrophysics Science Archive Research
Center (HEASARC), which is a service of the Astrophysics Science
Division at NASA/GSFC and the High Energy Astrophysics Division of the
Smithsonian Astrophysical Observatory;
the WEBDA database, operated at the
Institute for Astronomy of the University of Vienna;
and the Mikulski Archive for Space Telescopes (MAST). STScI is operated by the
Association of Universities for Research in Astronomy, Inc., under
NASA contract NAS5-26555. Support for MAST was provided by the NASA
Office of Space Science via grant NNX09AF08G and by other grants and
contracts.

Funding for SDSS-III has been provided by the Alfred P. Sloan
Foundation, the Participating Institutions, the National Science
Foundation, and the U.S. Department of Energy Office of Science. The
SDSS-III web site is http://www.sdss3.org/.

SDSS-III is managed by the Astrophysical Research Consortium for the
Participating Institutions of the SDSS-III Collaboration including the
University of Arizona, the Brazilian Participation Group, Brookhaven
National Laboratory, Carnegie Mellon University, University of
Florida, the French Participation Group, the German Participation
Group, Harvard University, the Instituto de Astrofisica de Canarias,
the Michigan State/Notre Dame/JINA Participation Group, Johns Hopkins
University, Lawrence Berkeley National Laboratory, Max Planck
Institute for Astrophysics, Max Planck Institute for Extraterrestrial
Physics, New Mexico State University, New York University, Ohio State
University, Pennsylvania State University, University of Portsmouth,
Princeton University, the Spanish Participation Group, University of
Tokyo, University of Utah, Vanderbilt University, University of
Virginia, University of Washington, and Yale University.

\facilities{MMT, FLWO: 1.5m,  TNG (HARPS-N), Sloan (APOGEE)}

\software{
  IRAF \citep{iraf1,iraf2},
  HARPS Data Reduction Software (v3.7), ELC \citep{elc}
}

\begin{longrotatetable}
\begin{deluxetable}{lc|ll|lll|lll|lll}
%\tablecolumns{13}
\tabletypesize{\scriptsize}
\tablewidth{0pc}
\tablecaption{Photometry of the Binary WOCS 11028 and Proxy Stars}
\tablehead{
& & & & \multicolumn{3}{|c}{Bright Component A} & \multicolumn{3}{|c}{Faint Component B} & \multicolumn{2}{|c}{Ratios} & \\
  & & \multicolumn{2}{|c}{WOCS 11028} & \multicolumn{2}{|c}{WOCS 6018} & \colhead{MS Fit} & \multicolumn{2}{|c}{WOCS 10027} & MS Fit & \colhead{Stars} & \colhead{MS Fit} & \\
\colhead{Filter} & $\lambda_{eff}$ (\AA) & \colhead{$m_\lambda$} & $\sigma_m$ & \colhead{$m_\lambda$} & \colhead{$\sigma_m$} & $m_\lambda$ & \colhead{$m_\lambda$} & \colhead{$\sigma_m$} & $m_\lambda$ & \colhead{$F_B / F_A$} & \colhead{$F_B / F_A$} & \colhead{Ref.}}
\startdata
$uvw2$ & 2030   & 16.715 & 0.058 & 16.898 & 0.040 & 16.886 &        &       & 19.389 &       & 0.100 &\\
$uvm2$ & 2231   & 16.655 & 0.072 & 16.757 & 0.050 &        &        &       &        &       &       &\\
NUV    & 2315.7 & 18.164 & 0.006 & 18.230 & 0.008 & 18.221 & 21.047 & 0.049 & 21.277 & 0.075 & 0.060 &\\% all calculated by combining all NUV observations
$uvw1$ & 2634   & 15.228 & 0.030 & 15.390 & 0.021 & 15.382 &        &       & 17.753 &       & 0.113 &\\
$U_V$  & 3450   &        &       & 15.953 & 0.018 &        &        &       & 18.030 &       & 0.145 & 1\\
$u$    & 3520   & 15.283 & 0.004 & 15.476 & 0.004 & 15.386 & 17.183 & 0.016 & 17.470 & 0.208 & 0.147 & 2\\
$u_{SDSS}$ & 3551 & 14.851 & 0.009 & 15.025 & 0.014 & 14.993 & 16.825 & 0.029 & 17.096 & 0.191 & 0.144 &\\
$U$    & 3663   &        &       & 14.144 &       & 14.202 & 15.962 &       & 16.308 & 0.187 & 0.144 & 3\\
$P_V$  & 3740   &        &       & 15.475 & 0.013 & 15.449 &       &       & 17.590 &       & 0.139 & 1\\
BATC2 ($b$) & 3890 & 14.39 & 0.01 & 14.59 & 0.02  & 14.539 & 16.45  & 0.05  & 16.695 & 0.180 & 0.137 & 4\\
$X_V$  & 4054   &        &       & 14.882 & 0.011 & 14.854 &       &       & 16.825 &        & 0.163 & 1\\
$v$    & 4100   & 14.308 & 0.003 & 14.522 & 0.002 & 14.466 & 16.155 & 0.014 & 16.401 & 0.222 & 0.168 & 2\\
$B$    & 4361   &        &       &        &       & 14.125 & 15.739 & 0.006 & 15.930 &       & 0.190 & 5\\%Sandquist; 1005 cal for S795
$B$    & 4361   &        &       & 14.149 &       & 14.149 & 15.758 & 0.015 & 15.960 & 0.227 & 0.189 & 3\\%MMJ
$B$    & 4361   & 13.971 & 0.068 & 14.164 & 0.031 & 14.118 & 15.786 & 0.096 & 15.955 & 0.224 & 0.184 & 6\\%APASS
$B$    & 4361   & 13.952 &       & 14.183 &       & 14.151 & 15.627 &       & 15.960 & 0.264 & 0.189 & 7 \\%Nardiello
$B$    & 4361   & 13.881 & 0.013 &        &       & 14.123 & 15.675 & 0.004 & 15.912 &       & 0.193 & 8 \\%Yadav
BATC4 ($d$) & 4532 & 13.67 & 0.01 & 13.92 & 0.01  & 13.879 & 15.41  & 0.01  & 15.620 & 0.254 & 0.201 & 4\\
$Y_V$  & 4665   &        &       & 14.148 & 0.008 & 14.117 &        &       & 15.827 &       & 0.207 & 1\\
$g^\prime$ & 4640    & 13.759 & 0.386 & 13.979 & 0.307 & 13.807 & 15.360 & 0.176 & 15.481 & 0.280 & 0.214 & 6 \\%APASS
$g_{SDSS}$ & 4686 & 13.592 & 0.001 & 13.825 & 0.008 & 13.816 & 15.292 & 0.010 & 15.491 & 0.259 & 0.214&\\
$b$    & 4688   & 13.720 & 0.002 & 13.979 & 0.002 & 13.940 & 15.454 & 0.011 & 15.653 & 0.257 & 0.207 & 2\\
$g_{P1}$ & 4810  & 13.560 & 0.001 & 13.784 & 0.001 & 13.743 & 15.224 & 0.003 & 15.425 & 0.266 & 0.212 &\\
BATC5 ($e$) & 4916 & 13.53 & 0.01 & 13.78 & 0.01  & 13.710 & 15.17  & 0.01  & 15.375 & 0.278 & 0.216 & 4\\
$G_{BP}$ & 5051.5 & 13.5113 & 0.0013 & 13.7714 & 0.0011 & 13.7289 & 15.1898 & 0.0016 & 15.3801 & 0.271 & 0.219 &\\
$Z_V$  & 5162   &        &       & 13.815 & 0.008 & 13.777 &        &       & 15.451 &       & 0.214 & 1\\
BATC6 ($f$) & 5258 & 13.38 & 0.01 & 13.65 & 0.01  & 13.585 & 15.03  & 0.02  & 15.225 & 0.281 & 0.221 & 4\\
$V_V$  & 5442   &        &       & 13.622 & 0.006 & 13.582 &        &       & 15.182 &       & 0.229 & 1\\
$V$    & 5448   & 13.337 & 0.001 & 13.597 & 0.001 & 13.564 & 14.953 & 0.002 & 15.152 & 0.287 & 0.232 & 5\\%1005 calibration for S617,1597
$V$    & 5448   &        &       & 13.598 &       & 13.597 & 15.010 & 0.010 & 15.183 & 0.275 & 0.227 & 3\\%MMJ
$V$    & 5448   & 13.366 & 0.042 & 13.619 & 0.011 & 13.566 & 14.977 & 0.047 & 15.158 & 0.286 & 0.231 & 6 \\%APASS
$V$    & 5448   & 13.391 &       & 13.675 &       & 13.597 &        &       & 15.174 &      & 0.234 & 7 \\%Nardiello
$V$    & 5448   & 13.246 & 0.004 &        &       & 13.539 & 14.924 & 0.007 & 15.123 &      & 0.232 & 8 \\%Yadav
$y$    & 5480   & 13.349 & 0.003 & 13.616 & 0.003 & 13.568 & 14.974 & 0.008 & 15.167 & 0.286 & 0.229 & 2\\
BATC7 ($g$) & 5785 & 13.24 & 0.02 & 13.55 & 0.02  & 13.473 & 14.84  & 0.03  & 15.013 & 0.305 & 0.242 & 4\\
BATC8 ($h$) & 6069 & 13.16 & 0.02 & 13.44 & 0.02  & 13.380 & 14.70  & 0.03  & 14.901 & 0.313 & 0.246 & 4\\
$r^\prime$ & 6122 & 13.194 & 0.019 & 13.491 & 0.019 & 13.440 & 14.737 & 0.041 & 14.952 & 0.317 & 0.248 & 6 \\%APASS
$r_{SDSS}$ & 6166 & 13.166 & 0.001 & 13.487 & 0.017 & 13.410 & 14.769 & 0.010 & 14.911 & 0.307 & 0.251 & \\
%$r_N$   & 6166  & 13.212 &       & 13.523 &       & 13.472 & 14.765 &       & 14.949 & 0.319 & 0.257 & 7\\%Nardiello re_cal
$r_{P1}$ & 6170  &          &     & 13.482 & 0.013 & 13.426 & 14.756 & 0.001 & 14.926 & 0.309 & 0.251 & \\ %bad S617 measurement
$G$    & 6230.6 & 13.1974 & 0.0002 & 13.4816 & 0.0002 & 13.4363 & 14.8044 & 0.0003 & 14.9717 & 0.296 & 0.243 &\\
$S_V$  & 6534   &        &       & 13.112 & 0.008 & 13.062 &        &       & 14.546 &       & 0.255 & 1\\
BATC9 ($i$) & 6646 & 13.13 & 0.02 & 13.44 & 0.01  & 13.394 & 14.66  & 0.03  & 14.859 & 0.325 & 0.259 & 4\\
BATC10a ($j$) & 7055 & 13.08 & 0.01 & 13.39 & 0.01 & 13.327 & 14.66 & 0.03  & 14.786 & 0.334 & 0.261 & 4\\
$i^\prime$ & 7440 & 13.264 & 0.369 & 13.528 & 0.305 & 13.374 & 14.545 & 0.072 & 14.790 & 0.392 & 0.271 & 6 \\%APASS
$i_{SDSS}$ & 7480 &       &       & 13.357 & 0.001 & 13.308 & 14.602 & 0.021 & 14.742 & 0.318 & 0.267 & \\
$i_{P1}$ & 7520  & 13.225 &       & 13.465 &       & 13.337 & 14.618 & 0.003 & 14.748 & 0.346 & 0.273 &\\
BATC10b ($k$) & 7545 & 13.06 & 0.01 & 13.37 & 0.01 & 13.308 & 14.53 & 0.04  & 14.740 & 0.344 & 0.267 & 4\\
$G_{RP}$ & 7726.2 & 12.7237 & 0.0007 & 13.0354 & 0.0006 & 12.9865 & 14.2558 & 0.0014 & 14.4074 & 0.325 & 0.270 & \\
$I_C$  & 7980   & 12.655 & 0.001 & 12.957 & 0.002 & 12.919 & 14.132 & 0.002 & 14.325 & 0.339 & 0.274 & 5\\% 1005 calibration for S1597
$I_C$  & 7980   &        &       & 12.920 &       & 12.888 & 14.176 &       & 14.325 & 0.314 & 0.266 & 3\\%MMJ
$I_C$  & 7900   &        &       & 12.981 & 0.008 & 12.934 &        &       & 14.355 &       & 0.270 & 1\\%Vilnius
$I_C$  & 7900   & 12.584 & 0.005 &        &       & 12.874 & 14.107 & 0.002 & 14.311 &       & 0.266 & 8\\%Yadav
BATC11 ($m$) & 8020 & 13.04 & 0.01 & 13.35 & 0.01 & 13.283 & 14.48  & 0.02  & 14.697 & 0.353 & 0.272 & 4\\
BATC12 ($n$) & 8483 & 13.01 & 0.01 & 13.34 & 0.01 & 13.289 & 14.49  & 0.01  & 14.679 & 0.347 & 0.278 & 4\\
$z_{P1}$ & 8660  & 13.058 & 0.004 & 13.387 & 0.003 & 13.339 & 14.579 & 0.002 & 14.715 & 0.334 & 0.282 &\\
$z_{SDSS}$ & 8932 & 13.010 & 0.016 & 13.350 & 0.011 & 13.291 & 14.602 & 0.021 & 14.661 & 0.344 & 0.283 &\\
BATC13 ($o$) & 9180 & 12.97 & 0.01 & 13.30 & 0.01 & 13.246 & 14.41  & 0.01  & 14.610 & 0.360 & 0.285 &\\
$y_{P1}$ & 9620  & 13.037 & 0.005 & 13.366 & 0.007 & 13.318 & 14.527 & 0.010 & 14.667 & 0.343 & 0.289 &\\
BATC14 ($p$) & 9736 & 12.99 & 0.01 & 13.31 & 0.01 & 13.281 & 14.43  & 0.01  & 14.630 & 0.360 & 0.289 &\\
$J$   & 12350   & 12.177 & 0.018 & 12.546 & 0.022 & 12.492 & 13.609 & 0.036 & 13.767 & 0.376 & 0.309 &\\
$H$   & 16620   & 11.903 & 0.020 & 12.260 & 0.020 & 12.237 & 13.276 & 0.042 & 13.391 & 0.392 & 0.346 &\\
$K_s$ & 21590   & 11.829 & 0.020 & 12.239 & 0.023 & 12.177 & 13.205 & 0.024 & 13.311 & 0.411 & 0.352 &\\
$W1$  & 33526   & 11.799 & 0.023 & 12.211 & 0.024 & 12.131 & 13.015 & 0.030 & 13.264 & 0.477 & 0.352 &\\
$W2$  & 46028   & 11.839 & 0.021 & 12.250 & 0.415 & 12.165 & 13.095 &       & 13.311 & 0.459 & 0.348 &\\
\hline
$T$ (K) &       &        &       & 6200   & 100   &  6200  & 5600   & 100   & 5500    & 0.88  & 0.89  &\\
$F_{bol}$ ($10^{-11}$ erg / cm$^2$ s) & & & & 10.47 & & 10.89 & 3.21 & & 2.73 & 0.310 & 0.248 & \\%for uncertainties 11.31,10.57; 3.15(5600),2.39(5400); ratio range: 0.211-0.298 (2sig)
$R$ ($\rsun$) & & & & & & & & & & 0.719 & 0.643 &\\
\enddata
\label{sedtable}
\tablerefs{1: \citet{laug}. 2: \citet{balaguer}. 3: \citet{mmj}. 4:
  \citet{zhoubatc}. 5: \citet{mephot}. 6: \citet{apass}. 7:
  \citet{nardiellom67}. 8: \citet{yadav}.}
\end{deluxetable}
\end{longrotatetable}

\clearpage
\begin{deluxetable*}{lcccccl}
\tablewidth{0pt}
\tablecaption{Detached Eclipsing Binary Stars with Masses Similar to WOCS 11028 A}
\tablehead{\colhead{Name} & \colhead{$M / \msun$} & \colhead{$E(B-V)$} & \colhead{$\omega$ (mas)} & \colhead{Filter} & \colhead{Magnitude} & \colhead{Refs.\tablenotemark{a}}
} 
\startdata
AD Boo B & $1.209\pm0.006$ & 0.034 & $5.118\pm0.073$ & $y$ & $10.721\pm0.013$ & 1 \\
         &                 &       &                 & $b$ & $11.092\pm0.007$ & \\
         &                 &       &                 & $v$ & $11.661\pm0.011$ & \\
         &                 &       &                 & $u$ & $12.570\pm0.016$ & \\ %Clausen et al. 2008 E(b-y)=0.025+/-0.015
BK Peg B & $1.257\pm0.005$ & 0.061 & $3.139\pm0.053$ & $y$ & $11.080\pm0.014$ & 2\\
         &                 &       &                 & $b$ & $11.440\pm0.008$ & \\
         &                 &       &                 & $v$ & $11.579\pm0.016$ & \\
         &                 &       &                 & $u$ & $12.514\pm0.017$ & \\ %Clausen et al. 2010 E(b-y)=0.044 
AP And A & $1.277\pm0.004$ & 0.058 & $2.866\pm0.041$ & $V$ & 11.845 & 3\\
AP And B & $1.251\pm0.004$ & 0.058 & $2.866\pm0.041$ & $V$ & 11.963 & 3\\%   Lacy et al. 2014  (on ZAMS)  [Fe/H]=-0.09 (phot)  E(B-V)=0.058+/-0.030
VZ Hya A & $1.271\pm0.009$ & 0.028 & $6.765\pm0.049$ & $y$ & $9.442\pm0.008$ & 1\\
         &                 &       &                 & $b$ & $9.723\pm0.005$ & \\
         &                 &       &                 & $v$ & $10.169\pm0.008$ & \\
         &                 &       &                 & $u$ & $11.035\pm0.009$ & \\%  Clausen et al. 2008 E(b-y)=0.020+/-0.020 [Fe/H]=-0.20+/-0.12
V501 Her A & $1.2690\pm0.0035$ & 0.048 & $2.309\pm0.022$ & $V$ & $11.623\pm0.048$ & 4\\ 
V501 Her B & $1.2113\pm0.0032$ & 0.048 & $2.309\pm0.022$ & $V$ & $12.188\pm0.048$ & 4\\%  Lacy & Fekel 2014  E(B-V)=0.048+/-0.018 (reddening maps)
EF Aqr A & $1.244\pm0.008$ & 0. & $5.765\pm0.060$ & $y$ & $10.015\pm0.022$ & 5\\
         &                 &    &                 & $b$ & $10.353\pm0.010$ & \\
         &                 &    &                 & $v$ & $10.874\pm0.010$ & \\
         &                 &    &                 & $u$ & $11.778\pm0.015$ & \\ % Vos et al. 2012 E(b-y)=0.000  [Fe/H]=0.00 (tabulated values are dereddeded for E(b-y)=0.018
WZ Oph A & $1.227\pm0.007$ & 0.045 & $6.372\pm0.033$ & $y$ & $9.856\pm0.017$ & 1\\
         &                 &       &                 & $b$ & $10.218\pm0.005$ & \\
         &                 &       &                 & $v$ & $10.729\pm0.008$ & \\
         &                 &       &                 & $u$ & $11.609\pm0.011$ & \\ % Clausen et al. 2008 E(b-y)=0.033+/-0.012  [Fe/H]=-0.27+/-0.07
WZ Oph B & $1.220\pm0.006$ & 0.045 & $6.372\pm0.033$ & $y$ & $9.841\pm0.017$ & 1\\
         &                 &       &                 & $b$ & $10.211\pm0.005$ & \\
         &                 &       &                 & $v$ & $10.722\pm0.008$ & \\
         &                 &       &                 & $u$ & $11.615\pm0.012$ & \\
UX Men A & $1.2229\pm0.0015$ & 0.028 & $9.644\pm0.025$ & $y$ & $8.89\pm0.01$ & 6\\
         &                 &       &                 & $b$ & $9.249\pm0.005$ & \\
         &                 &       &                 & $v$ & $9.769\pm0.005$ & \\
         &                 &       &                 & $u$ & $10.660\pm0.005$ & \\
UX Men B & $1.1878\pm0.0015$ & 0.028 & $9.644\pm0.025$ & $y$ & $9.07\pm0.01$ & 6\\
         &                 &       &                 & $b$ & $9.438\pm0.005$ & \\
         &                 &       &                 & $v$ & $9.990\pm0.005$ & \\
         &                 &       &                 & $u$ & $10.899\pm0.005$ & \\ %   Helminiak et al. 2009 (Andersen et al. 1989 phot) E(b-y)=0.02 [Fe/H]=0.04  ~2.7Gyr
FL Lyr A & $1.2102\pm0.0076$ & 0.007 & $7.406\pm0.025$ & $V$ & $9.58$ & 7 \\
% $R_C$ & $9.13\pm0.01$ & \\ %Helminiak et al. 2019 (Popper 1986 gives Lp/Ltot)E(b-y)=0.005 [Fe/H]=-0.07
LL Aqr A & $1.1959\pm0.0007$ & 0.018 & $7.269\pm0.051$ & $V$ & $9.71\pm0.012$ & 8\\
         &                   &       &                 & $B$ & $10.224\pm0.032$ & \\
         &                   &       &                 & $U$ & $10.160\pm0.062$ & \\ % Graczyk et al. 2016 [Fe/H]=0.02 E(B-V)=0.018
WOCS 40007 A & $1.218\pm0.008$ & 0.16 & 0.420\tablenotemark{b} & $V$ & $16.053\pm0.02$ & 9\\
             &                 &      &       & $B$ & $16.669\pm0.006$ & \\
             &                 &      &       & $I_C$ & $15.307\pm0.025$ & \\ % Jeffries et al. 2013 [Fe/H]=-0.03  E(B-V)=0.16+/-0.007 (from spreadsheet with correction for differential reddening,tertiary; "parallax" based on DM, not Gaia)
\enddata
\label{debmags}
\tablenotetext{a}{References: 1: \citet{clausen08}. 2: \citet{clausen10}.
  3: \citet{lacy14}. 4: \citet{lf14}. 5: \citet{vos12}. 6: \citet{helm09,acm89}. 7: \citet{helm19,popper86}. 8: \citet{grac16}. 9: \citet{brewer}.}
\tablenotetext{b}{Distance modulus from \citet{brewer} translated back into an equivalent parallax.}
\end{deluxetable*}

\end{document}